\documentclass[pra,aps,showpacs,groupedaddress,12pt,nofootinbib]{revtex4}
\usepackage{amssymb,amsmath,amstext,amsfonts,mathrsfs,bm,accents,mathtools,enumerate}

\begin{document}
 \title{Could the Universe have an Exotic Topology?}
 \author{Vincent Moncrief}
 \affiliation{Department of Physics and Department of Mathematics, \\ Yale University, P.O. Box 208120, New Haven, CT 06520, USA. \\ E-mail address: vincent.moncrief@yale.edu}
 \author{Puskar Mondal}
 \affiliation{\hspace{50pt} Department of Geophysics, \hspace{50pt} \\ Yale University, 210 Whitney Avenue, New Haven, CT 06511 \\ P.O. Box 208109, New Haven, CT 06520-8109, USA. \\ E-mail address: puskar.mondal@yale.edu}
 \begin{abstract}
 A recent article uncovered a surprising dynamical mechanism at work within the (vacuum) Einstein `flow' that strongly suggests that many closed 3-manifolds that do not admit a locally homogeneous and isotropic metric \textit{at all} will nevertheless evolve, under Einsteinian evolution, in such a way as to be \textit{asymptotically} compatible with the observed, approximate, spatial homogeneity and isotropy of the universe \cite{Moncrief:2015}. Since this previous article, however, ignored the potential influence of \textit{dark-energy} and its correspondent accelerated expansion upon the conclusions drawn, we analyze herein the modifications to the foregoing argument necessitated by the inclusion of a \textit{positive} cosmological constant --- the simplest viable model for dark energy.
 \end{abstract}
 \maketitle

\section{Introduction and Topological Background}
 \label{sec:topological}

 Viewed on a sufficiently coarse-grained scale the portion of our universe that is accessible to observation appears to be spatially homogeneous and isotropic. If, as is usually imagined, one should be able to extrapolate these features to (a suitably coarse-grained model of) the universe as a whole then only a handful of spatial manifolds need be considered in cosmology --- the familiar Friedmann-Lema\^{i}tre-Robertson-Walker (FLRW) archetypes of constant positive, vanishing or negative curvature \cite{Nussbaumer:2009, Ellis:2012}. These geometries consist, up to an overall, time-dependent scale factor, of the 3-sphere, \(\mathbb{S}^{3}\), with its canonical `round' metric, Euclidean 3-space, \(\mathbb{E}^{3}\), hyperbolic 3-space, \(\mathbb{H}^{3}\) and the quotient space \(\mathbb{R}P(3) \approx \mathbb{S}^{3}/\pm I\) obtainable from \(\mathbb{S}^{3}\) by the identification of antipodal points \cite{Wolf:2010}. Of these possibilities only the sphere and its 2-fold quotent \(\mathbb{R}P(3)\) are \textit{closed} and thus compatible with a universe model of finite extent. It is not known of course whether the actual universe is spatially closed or not but, to simplify the present discussion, we shall limit our attention herein to models that are. More precisely we shall focus on spacetimes admitting Cauchy hypersurfaces that are each diffeomorphic to a smooth, connected 3-manifold that is compact, orientable and without boundary.

 On the other hand if one takes literally the cosmological principle that only manifolds supporting a \textit{globally} homogeneous and isotropic metric should be considered in models for the actual universe then, within the spatially compact setting considered here, only the 3-sphere and \(\mathbb{R}P(3)\) would remain. But the astronomical observations which motivate this principle are necessarily limited to a (possibly quite small) fraction of the entire universe and are compatible with models admitting metrics that are only locally, but not necessarily globally, spatially homogeneous and isotropic. As is well-known there are spatially compact variants of all of the basic Friedmann-Lema\^{i}tre-Robertson-Walker cosmological models, mathematically constructable (in the cases of vanishing or negative curvature) by taking suitable compact quotients of Euclidean 3-space $\mathbb{E}^3$ or of hyperbolic 3-space \(\mathbb{H}^3\). One can also take infinitely many possible quotients of $\mathbb{S}^3$ to obtain the so-called \textit{spherical space forms} that are locally compatible with the FLRW constant positive curvature geometry but are no longer diffeomorphic to the 3-sphere.

	Still more generally though we shall find that there is a dynamical mechanism at work within the Einstein `flow', suitably viewed in terms of the evolution of  3-manifolds to develop 4-dimensional, globally hyperbolic spacetimes, and extended to include a positive cosmological constant \(\Lambda\), that strongly suggests that even manifolds that do not admit a locally homogeneous and isotropic metric \textit{at all} will nevertheless evolve in such a way as to be \textit{asymptotically} compatible with the observed homogeniety and isotropy. This reflects an argument which we shall sketch that, under Einsteinian-\(\Lambda\) evolution, the summands making up M (in a connected sum decomposition) that do support locally homogeneous and isotropic metrics will tend to overwhelmingly dominate the spatial volume asymptotically as the universe model continues to expand and furthermore that the actual evolving (inhomogeneous, non-isotropic) metric on M will naturally tend to flow towards a homogeneous, isotropic one on each of these asymptotically volume-dominating summands.

	We do not claim that this mechanism is yet so compelling, either mathematically or physically, as to convince one that the actual universe has a more exotic topology but only that such a possibility is not strictly excluded by current observations. However, it is intriguing to investigate the possibility that there may be a dynamical reason, provided by Einstein's equations, for the observed fact that the universe seems to be at least locally homogeneous and isotropic and that this mechanism may therefore allow an attractive logical alternative to simply extrapolatating observations of necessarily limited scope to the universe as a whole.

	But what are the (compact, connected, orientable) 3-manifolds available for consideration? This question has been profoundly clarified in recent years by the dramatic progress on lower dimensional topology made possible through the advancements in Ricci flow \cite{ricci-flow}. One now knows for example that, since the Poincar\'{e} conjecture has finally been proven, any such 3-manifold \(M\) that is in fact \textit{simply connected} must be diffeomorphic to the ordinary 3-sphere \(S^{3}\). Setting aside this so-called `trivial' manifold the remaining possibilities consist of an infinite list of nontrivial manifolds, each of which is diffeomorphic (designated herein by \(\approx\)) to a finite connected sum of the following form:
\begin{equation}
\begin{split}
&M \approx \\
& \underbrace{S^3/\Gamma_1 \;\# \cdots \#\; S^3/\Gamma_k}_{k\; \hbox{spherical factors}} \;\#\; \underbrace{(S^2 \times S^1)_1\; \# \cdots \# (S^2 \times S^1)_\ell}_{\ell\; \hbox{wormhols (or handles)}}\; \#\; \underbrace{K(\pi_1,1)_1\; \# \cdots \#\; K(\pi_1,1)_m}_{m\; \hbox{aspherical factors}}
\end{split}
\label{eq:01}
\end{equation}
Here $k$, $\ell$ and $m$ are integers $\geq 0$, $k + \ell + m \geq 1$ and if either $k$, $\ell$  or $m$ is 0 then terms of that type do not occur. The connected sum $M\; \#\; N$ of two closed connected, oriented  n-manifolds is constructed by removing the interiors of an embedded closed n-ball in each of M and N and then identifying the resulting $S^{n - 1}$ boundary components by an orientation-reversing diffeomorphism. The resulting n-manifold will be smooth, connected, closed and consistently oriented with the original orientations of M and N. The above decomposition of M is only uniquely defined provided we set aside $S^3$ since  $M' \; \#\;  S^3  \approx M'$ for any 3-manifold $M'$.

	In the above formula if $k \geq 1$, then each $\Gamma_i , 1 \leq i \leq k$ is a finite, nontrivial ($\Gamma_i \neq [I]$) subgroup of SO(4) acting freely and orthogonally on  $\mathbb{S}^3$. The individual summands $S^3 /\Gamma_i$  are the \textit{spherical space forms} alluded to previously and, by construction, each is compatible with an FLRW metric of constant positive spatial curvature (i.e., $\mathrm{k} = +1$ models in the usual notation). The individual `handle' summands  $S^2 \times S^1$  admit metrics of the Kantowski-Sachs type that are homogeneous but not isotropic and so not even locally of FLRW type.

	The remaining summands in the above `prime decomposition' theorem \cite{Kneser:1929,Milnor:1962,Scott:1983} are the $K(\pi, 1)$ manifolds of Eilenberg-MacLane type wherein, by definition  $\pi =  \pi_1 (M)$, the fundamental group of M and all of the higher homotopy groups are trivial, that is $\pi_i (M) = 0 \hbox{ for } i > 1$. Equivalently, the universal covering space of M is contractible and, in this case, known to be diffeomorphic to $\mathbb{R}^3$ \cite{ricci-flow-b}. Since the higher homotopy groups,  $\pi_i (M)$ for $i > 1$, can be interpreted as the homotopy classes of continuous maps  $S^i \rightarrow M$, each such map must be homotopic to a constant map. For this reason $K(\pi, 1)$ manifolds are said to be aspherical.

	This general class of $K(\pi, 1)$ manifolds includes, as special cases, the 3-torus and five additional manifolds, finitely covered by the torus, that are said to be of `flat type' since they are the only compact, connected, orientable 3-manifolds that each, individually, admits a flat metric and thus supports spatially compactified versions of the FLRW spaces of flat type (i.e., $\mathrm{k} = 0$ models).

	Other $K(\pi, 1)$ spaces include the vast set of compact hyperbolic manifolds $\mathbb{H}^3 /\Gamma$ where here $\Gamma$ is a discrete torsion-free (i.e., no nontrivial element has finite order) co-compact subgroup of the Lie group Isom$^+$ ($\mathbb{H}^3$) of orientation-preserving isometries of $\mathbb{H}^3$  that, in fact, is Lie-group isomorphic to the proper orthochronous Lorentz group $SO^\dagger (3, 1)$. Each of these, individually, supports spatially compactified versions of the FLRW spacetimes of constant negative (spatial) curvature (i.e., $\mathrm{k} = -1$ models).

	Additional $K(\pi, 1)$ manifolds include the trivial circle bundles over higher genus surfaces $\Sigma_p$  for $p \geq 2$ (where $\Sigma_p$ designates a compact, connected, orientable surface of genus $p$) and nontrivial circle bundles over $\Sigma_p$  for $p \geq 1$. Note that the trivial circle bundles  $S^2 \times S^1$ and  $T^2 \times S^1 \approx T^3$ are already included among the previous prime factors discussed  and that nontrivial circle bundles over $S^2$ are included among the spherical space forms $S^3 /\Gamma$ for suitable  choices of $\Gamma$. The circle bundles over higher genus surfaces will reappear later as the basic spatial manifolds occurring in the so-called \(U(1)\) problem. Still further examples of $K(\pi, 1)$ manifolds are compact 3-manifolds that fiber nontrivially over the circle with fiber  $\Sigma_p$  for $p \geq 1$. Any such manifold is obtained by identifying the boundary components of $[0, 1] \times \Sigma_p$ with a (nontrivial) orientation-reversing diffeomorphism of  $\Sigma_p$.

	It is known however that every prime $K(\pi, 1)$ manifold is decomposible into a (possibly trivial but always finite) collection of (complete, finite volume) hyperbolic and \textit{graph manifold} components. The possibility of such a (nontrivial) decomposition arises whenever the $K(\pi, 1)$ manifold under study admits a nonempty family $\{T_i\}$ of disjoint embedded incompressible two-tori. An embedded two-torus $T^2$ is said to be incompressible if every incontractible loop in the torus remains incontractible when viewed as a loop in the ambient manifold. A closed oriented 3-manifold G (possibly with boundary) is a \textit{graph manifold} if there exists a finite collection $\{T'_i\}$ of disjoint embedded incompressible tori $T'_i \subset G$ such that each component $G_j$  of $G\; \backslash \cup T'_i$  is a Seifert-fibered space.\footnote{A Seifert-fibered space is a 3-manifold foliated by circular fibers in such a way that each fiber has a tubular neighborhood (characterized by a pair of co-prime integers) of the special type known as a standard fibered torus.\label{note01}} Thus a graph manifold is a union of Seifert-fibered spaces glued together by toral automorphisms along toral boundary components. The collection of tori is allowed to be empty so that, in particular, a Seifert-fibered manifold itself is a graph manifold. Decomposing a 3-manifold by cutting along essential two-spheres (to yield its prime factors) and then along incompressible tori, when present, are the basic operations that reduce a manifold to its `geometric' constituents \cite{Scott:1983}. The Thurston conjecture that every such 3-manifold can be reduced in this way has now been established via arguments employing Ricci flow \cite{ricci-flow}.

	For comparison's sake we recall that the two dimensional analogue of the foregoing (prime) decomposition theorem is the classical result that any compact, connected, orientable surface is either  $S^2$,  $T^2$ or a higher genus surface $\Sigma_p$  diffeomorphic to the connected sum of $p$ 2-tori for  $p \geq 2$. These surfaces provide the spatial topologies for `cosmological' $2 + 1$ dimensional Einstein gravity and, as we have mentioned, circle bundles over these provide the arenas for the U(1) problem in full 3 + 1 dimensional gravity.

	It may seem entirely academic to consider such general, `exotic' 3-manifolds as the composite (i.e., nontrivial connected sum) ones described above as arenas for general relativity when essentially all of the explicitly known solutions of Einstein's equations (in this spatially compact setting) involve only individual, `prime factors'. As we shall see however some rather general conclusions are derivable concerning the behaviors of solutions to the field equations on such exotic manifolds and astronomical observations do not logically exclude the possibility that the actual universe could have such a global topological structure. It is furthermore conceivable that the validity of central open issues in general relativity like the \textit{cosmic censorship conjecture} could depend crucially upon the spatial topology of the spacetime under study.

\section{Yamabe Classification}
\label{sec:Yamabe}

	In the following we shall focus attention on the subset of these 3-manifolds of so-called \textit{negative Yamabe type}. By definition these admit no Riemannian metric $\gamma$ having scalar curvature $R(\gamma)  \geq  0$. Within the above setting a closed 3-manifold M is of negative Yamabe type if and only if it lies in one of the following three mutually exclusive subsets \cite{Fischer:2006}:
\begin{enumerate}
\item M is hyperbolizable (that is admits a hyperbolic metric);
\item M is a non-hyperbolizable $K(\pi, 1)$ manifold of non-flat type
(the six flat $K(\pi, 1)$ manifolds are of zero Yamabe type);
\item M has a nontrivial connected sum decomposition (i.e., M is composite)
in which at least one factor is a $K(\pi, 1)$ manifold. In this case the $K(\pi, 1)$ factor may be either of flat type or hyperbolizable.
\end{enumerate}
The six flat manifolds comprise by themselves the subset of zero Yamabe type. These admit metrics having vanishing scalar curvature (the flat ones) but no metrics having strictly positive scalar curvature. Finally manifolds of positive Yamabe type provide the complement to the above two sets and include the stand-alone $S^3$, the spherical space forms $S^3 /\Gamma_i , S^2 \times S^1$  and connected sums of the latter two types (recalling that $M' \# S^3  \approx M'$ for any 3-manifold $M'$).

	It follows immediately from the form of the Hamiltonian constraint that any solution of the Einstein field equations with Cauchy surfaces of negative Yamabe type (i.e., diffeomorphic to a manifold in one of the three subsets listed above) and strictly non-negative energy density and non-negative cosmological constant (with either or both allowed to vanish) cannot admit a maximal hypersurface. Thus such a universe model, if initially expanding, can only continue to do so (until perhaps developing a singularity) and cannot cease its expansion and `recollapse'.

	For such manifolds Yamabe's theorem \cite{Lee:1987} guarantees that each smooth Riemannian metric on $M$ is uniquely, globally conformal to a metric $\gamma$ having scalar curvature $R(\gamma) = -1$. Thus, in a suitable function space setting \cite{Fischer:1975}, one can represent the conformal classes of Riemannian metrics on M by the infinite dimensional submanifold:
\begin{equation}
\mathcal{M}_{-1} (M) =  \{ \gamma \in  \mathcal{M} (M) | R(\gamma) = -1\}
\label{eq:02}
\end{equation}
where $\mathcal{M} (M)$ designates the corresponding space of arbitrary Riemannian metrics on M.

	The quotient of $\mathcal{M}_{-1} (M)$ by the natural action of $\mathcal{D}_0 (M) = \mathrm{Diff}_0(M)$, the connected component of the identity of the group $\mathcal{D}^+ (M) = \mathrm{Diff}^+ (M)$ of smooth, orientation preserving diffeomorphisms of M, defines an orbit space (not necessarily a manifold) given by   $\mathfrak{T}(M) =  \mathcal{M}_{-1} (M) / \mathcal{D}_0 (M)$. Because of it's resemblance to the corresponding Riemannian construction of the actual Teichm\"{u}ller space $\mathfrak{T}  (\Sigma_p )$ for a higher genus surface  $\Sigma_p$  \cite{Tromba:1992} we refer to  $\mathfrak{T} (M)$ (informally) as the `Teichm\"{u}ller space of conformal structures' of M. The actual Teichm\"{u}ller space $\mathfrak{T}  (\Sigma_p )$ of the higher genus surface $\Sigma_p$ is diffeomorphic to $\mathbb{R}^{6p - 6}$  hence always a smooth manifold. By contrast   $\mathfrak{T} (M)$ may either be a manifold or have orbifold singularities or consist of a stratified union of manifolds representing the different isometry classes of conformal Riemannian metrics admitted by M (i.e., metrics $\gamma$ with  $R(\gamma) = -1)$.

	In certain cases however, $\mathfrak{T}  (M)$ proves to be a global, smooth and even contractible (infinite-dimensional) manifold \cite{Fischer:1996} and thus to have all of the essential features (except the finite dimensionality) of an actual Teichm\"{u}ller space. The infinite dimensionality of this Teichm\"{u}ller-like space, $\mathfrak{T}  (M)$, which will play the role of reduced configuration space for Einstein's equations (in the vacuum case for simplicity), is of course needed to accommodate the gravitational wave degrees of freedom that are absent in $2 + 1$ dimensional Einstein gravity \cite{Moncrief:1989,Moncrief:2008}. One could perhaps however argue that a still more natural choice for the reduced configuration space would be the analogue of Riemann moduli space, wherein one would quotient $\mathcal{M}_{-1} (M)$ by the full group, $\mathcal{D}^+ (M) = \mathrm{Diff}^+ (M)$ of orientation preserving diffeomorphisms of M, instead of just its identity component. But since this construction invariably introduces orbifold singularities even in the $2 + 1$ dimensional problem it would also disturb the smooth character of even the favorable cases mentioned above in $3 + 1$ dimensions. For this reason we shall retain  $\mathfrak{T} (M)$ as our preferred definition for the reduced configuration space keeping in mind that the different conformal classes of M may thus not be uniquely represented.

\section{The Gauge Fixed Einstein-\(\Lambda\) Field Equations}
\label{sec:gauge-fixed-Einstein}
Let \textit{M} be a compact, connected, orientable \(C^\infty\) manifold of dimension \(n \geq 2\) and set \(\bar{M} = R \times M\). We define \(t:\bar{M} \rightarrow R\) by projection onto the first factor and consider Lorentzian metrics \(\bar{g}\) on \(\bar{M}\) such that the level sets of \(t, M_t = \lbrace t\rbrace \times M\), are Cauchy hypersurfaces of the spacetime \(\lbrace\bar{M},\bar{g}\rbrace\). When there is no chance for confusion we shall simply  write \textit{M} for \(M_t\).

Given \(\lbrace\bar{M},\bar{g}\rbrace\) let \textit{T} be the unit (timelike) future directed normal field to \(M_t\) and define the \textit{lapse} function \textit{N} and \textit{shift} vector field \textit{X} such that
\begin{equation}\label{eq:100}
\partial_t = NT + X
\end{equation}
with \(N > 0\) (since, by assumption \textit{T} is future directed) and \textit{X} tangent to \(M_t\).

Letting \(\lbrace x^i\rbrace_{i = 1}^n\) be local charts for \textit{M} and taking \(\lbrace x^\alpha\rbrace_{\alpha = 0}^n = \lbrace t, x^1, \ldots , x^n\rbrace\), with \(x^0 = t\), as corresponding (local) charts for \(\bar{M}\) we can express the Lorentz metric \(\bar{g}\) in the form
\begin{equation}\label{eq:101}
\bar{g} = -N^2 dt \otimes dt + g_{ij} (dx^i + X^i dt) \otimes (dx^j + X^j dt)
\end{equation}
where \(g := g_{ij} dx^i \otimes dx^j\) is the induced, Riemanian metric on \textit{M}. The second fundamental form \textit{k} of \textit{M} in \(\bar{M}\) is given by
\begin{equation}\label{eq:102}
k_{ij} = -\frac{1}{2N} \left( \partial_t g_{ij} - (\mathcal{L}_X g)_{ij}\right)
\end{equation}
where \(\mathcal{L}\) denotes the Lie derivative operator. The corresponding mean curvature \(\tau\) of \textit{M} is given by
\begin{equation}\label{eq:103}
\tau := \operatorname{tr}_g k = g^{ij} k_{ij}
\end{equation}
where
\begin{equation}\label{eq:104}
g^{ij} \frac{\partial}{\partial x^i} \otimes \frac{\partial}{\partial x^j} = g^{-1}
\end{equation}
is the contravariant form of the metric \textit{g} and \(\operatorname{tr}_g\) denotes `trace' with respect to \textit{g}.

The vacuum Einstein equations with a cosmological constant \(\Lambda\) (referred to herein as the Einstein-\(\Lambda\) equations) given by
\begin{equation}\label{eq:105}
\bar{R}_{\mu\nu} (\bar{g}) - \frac{1}{2} \bar{R} (\bar{g}) \bar{g}_{\mu\nu} + \Lambda \bar{g}_{\mu\nu} = 0
\end{equation}
can be written as a system of evolution and constraint equations for \textit{(g,k)}. The Einstein-\(\Lambda\) evolution equations are
\begin{subequations}\label{eq:106}
\begin{align}
\partial_t g_{ij} &= -2N k_{ij} + (\mathcal{L}_X g)_{ij},\label{eq:106a}\\
\partial_t k_{ij} &= -\nabla_i \nabla_j N + N \left(R_{ij} {(g)} + (\operatorname{tr}_g k) k_{ij} - 2k_{im} k_{\hphantom{m}j}^m - \frac{2\Lambda}{n - 1} g_{ij}\right) + (\mathcal{L}_X k)_{ij}\label{eq:106b}
\end{align}
\end{subequations}
whereas the constraints take the form
\begin{subequations}\label{eq:107}
\begin{align}
 R(g) - |k|^2 + (\operatorname{tr}_g k)^2 &= 2\Lambda,\label{eq:107a} \\
 \nabla^j k_{ij} - \nabla_i (\operatorname{tr}_g k) &= 0, \label{eq:107b}
\end{align}
\end{subequations}
where \(\nabla_i\) designates covariant differentiation with respect to \textit{g}, \(|k|^2 = k_{ij} k^{ij}\) and where \(R_{ij}(g)\) and \(R(g)\) are the Ricci tensor and curvature scalar of this metric.

A solution to the Einstein-\(\Lambda\) evolution and constraint equations is a curve \(t \mapsto (g, k, N, X)\) which satisfies (\ref{eq:106}--\ref{eq:107}). Assuming sufficient regularity the spacetime metric \(\bar{g}\) given in terms of \((g, N, X)\) by (\ref{eq:101}) solves the vacuum Einstein-\(\Lambda\) field equations (\ref{eq:105}) if and only if the corresponding curve \((g, k, N, X)\) solves (\ref{eq:106}--\ref{eq:107}). The system (\ref{eq:106}--\ref{eq:107}) is however not hyperbolic so that to get a well-posed evolution problem we must modify this system by suitably fixing the gauge.

Let \(\hat{g}\) be a fixed \(C^\infty\) Riemannian metric on \textit{M} with Levi-Civita covariant derivative \(\hat{\nabla}\) and Christoffel symbols \(\hat{\Gamma}_{ij}^k\). Define the vector field \(V^k\) by
\begin{equation}\label{eq:108}
V^k = g^{ij} \left(\Gamma_{ij}^k(g) - \hat{\Gamma}_{ij}^k(\hat{g})\right).
\end{equation}
Then \(-V^k\) is the `tension field' of the identity map \(Id:(M,g) \rightarrow (M,\hat{g})\) so that \textit{Id} is harmonic precisely when \(V^k = 0\) (see \cite{Eells:1978} for background on harmonic maps).

The constant mean curvature and spatial harmonic (CMCSH) gauge condition that we shall employ is defined by the equations
\begin{subequations}\label{eq:109}
\begin{align}
  \operatorname{tr}_g k  &:= \tau = t\quad \hbox{(constant mean curvature)}\label{eq:109a} \\
  V^k &= 0\quad \hbox{(spatial harmonic coordinates)} \label{eq:109b}
\end{align}
\end{subequations}

Let the second order operator \(\hat{\Delta}_g\) be defined on symmetric 2-tensors by
\begin{equation}\label{eq:110}
(\hat{\Delta}_g h)_{ij} = \frac{1}{\mu_g} \hat{\nabla}_m \left(g^{mn} \mu_g (\hat{\nabla}_n h)_{ij}\right)
\end{equation}
where \(\mu_g = \sqrt{\det{g}}\) is the volume element on \((M,g)\). Using the identity \(\hat{\nabla}_m (\mu_g g^{-1})^{mn} = -V^n \mu_g,\; (\hat{\Delta}_g h)_{ij}\) may be written in the form
\begin{equation}\label{eq:111}
(\hat{\Delta}_g h)_{ij} = g^{mn} (\hat{\nabla}_m \hat{\nabla}_n h)_{ij} - V^m (\hat{\nabla}_m h)_{ij}
\end{equation}
so that, if the gauge condition \(V^k = 0\) is satisfied, \((\hat{\Delta}_g h)_{ij} \rightarrow g^{mn} (\hat{\nabla}_m \hat{\nabla}_n h)_{ij}\). A computation shows that
\begin{equation}\label{eq:112}
R_{ij}(g) = -\frac{1}{2} (\hat{\Delta}_g g)_{ij} + \mathcal{S}_{ij} [g, \partial g] + \alpha_{ij}
\end{equation}
where the symmetric tensor \(\alpha_{ij}\) is defined by
\begin{equation}\label{eq:113}
\alpha_{ij} = \frac{1}{2} (\nabla_i V_j + \nabla_j V_i)
\end{equation}
and \(\mathcal{S} [g,\partial g]\) is at most of quadratic order in the first derivatives of \(g_{ij}\) (c.f. Section 3 of Ref. \cite{Andersson:2003} for the explicit formula). Thus the system \(g_{ij} \mapsto R_{ij}(g) - \alpha_{ij}\) is quasi-linear elliptic.

In order to construct solutions to the Cauchy problem for the system consisting of the Einstein-\(\Lambda\) evolution and constraint equations (\ref{eq:106}--\ref{eq:107}) together with the gauge conditions (\ref{eq:109}) we shall consider the following modified form of the Einstein-\(\Lambda\) evolution equations
\begin{subequations}\label{eq:114}
\begin{align}
  \partial_t g_{ij} &= -2N k_{ij} + (\mathcal{L}_X g)_{ij}, \label{eq:114a} \\
  \partial_t k_{ij} &= -\nabla_i \nabla_j N + N \left(R_{ij}(g) + (\operatorname{tr}_g k) k_{ij} -2k_{im} k_{\hphantom{m}j}^m - \frac{2\Lambda}{n-1} g_{ij} - \alpha_{ij}\right) + (\mathcal{L}_X k)_{ij} \label{eq:114b}
\end{align}
\end{subequations}
coupled to the elliptic defining equations for \textit{N, X}, needed to preserve the imposed gauge conditions,
\begin{subequations}\label{eq:115}
 \begin{align}
 -\Delta_g N + N \left(|k|^2 - \frac{2\Lambda}{n-1}\right) &=1,\label{eq:115a}\\
 \begin{split}
 (\Delta_g X)^i + R_{\hphantom{i}m}^i(g) X^m - (\mathcal{L}_X V)^i &= (-2N k^{mn} + 2\nabla^m X^n) \left(\Gamma_{mn}^i(g) - \hat{\Gamma}_{mn}^i(\hat{g})\right)\\
  &\hphantom{=} {} + 2 (\nabla^m N) k_{\hphantom{i}m}^i - (\nabla^i N) k_{\hphantom{m}m}^m\end{split}\label{eq:115b}
 \end{align}
\end{subequations}
where \((\Delta_g X)^i := (\nabla^m \nabla_m X)^i, \Delta_g N := \nabla^m \nabla_m N,\), etc.

If \(V^k = 0\), so that \(\alpha_{ij} = 0\), then (\ref{eq:114}) coincides with the Einstein-\(\Lambda\) evolution equations specialized to the spatial harmonic (SH) gauge. In the Appendix below we shall sketch the proof that the system (\ref{eq:114}--\ref{eq:115}), supplemented by the gauge conditions (\ref{eq:109}) satisfies a local well-posedness theorem so that, in particular, the constraints and gauge conditions are conserved by the evolution equations. It will thus follow that the original system (\ref{eq:106}--\ref{eq:107}) obeys a well-posed Cauchy problem when specialized to the chosen (CMCSH) gauge. In view of the fact that \(g_{ij} \mapsto R_{ij}(g) - \alpha_{ij}\) is elliptic, the system (\ref{eq:114}) is hyperbolic and the coupled system (\ref{eq:114}--\ref{eq:115}) is elliptic-hyperbolic.

Whereas our proof of well-posedness is, for convenience, expressed in terms of the `Lagrangian' variables \((g, k)\) it will be useful, for the balance of our analysis, to convert the evolution, constraint and gauge equations to the corresponding Hamiltonian picture by reexpressing them in terms of `canonical' variables \((g, \pi)\) where the momentum \(\pi\), conjugate to the first fundamental form \textit{g}, is defined, via a Legendre transformation, by
\begin{equation}\label{eq:116}
\pi^{ij} = -\mu_g \left(k^{ij} - (\operatorname{tr}_g k) g^{ij}\right).
\end{equation}

The Hamiltonian dynamics of the Einstein-\(\Lambda\) system is given by the following complete set of evolution and constraint equations
\begin{subequations}\label{eq:117}
 \begin{align}
 \partial_t g_{ij} &= \frac{2N}{\mu_g} \left(\pi_{ij} - \left(\frac{\operatorname{tr}_g \pi}{n-1}\right) g_{ij}\right) + \nabla_i X_j + \nabla_j X_i,\label{eq:117a}\\
 \begin{split}
 \partial_t \pi^{ij} &= -N\mu_g \left(R^{ij}(g) - \frac{1}{2} g^{ij} R(g)\right) + \frac{N g^{ij}}{2\; \mu_g} \left(\pi^{mn} \pi_{mn} - \frac{(\operatorname{tr}_g \pi)^2}{n-1}\right)\\
 & \hphantom{=} {} -\frac{2N}{\mu_g} \left(\pi^{im} \pi_{\hphantom{j}m}^{j} - \frac{\pi^{ij}}{n-1} (\operatorname{tr}_g \pi)\right) + \mu_g (\nabla^i \nabla^j N - g^{ij} \nabla^k \nabla_k N)\\
 & \hphantom{=} {} -N \Lambda \mu_g g^{ij} + \nabla_m (X^m \pi^{ij}) - (\nabla_m X^i) \pi^{mj} - (\nabla_m X^j) \pi^{im}, \end{split}\label{eq:117b}\\
 \mathcal{J}_i(g,\pi) &:= -2\; \nabla_j \pi_{\hphantom{j}i}^j = 0,\label{eq:117c}\\
 \begin{split}
 \mathcal{H}(g, \pi) &:= \frac{1}{\mu_g} \left(\pi_{\hphantom{j}i}^j \pi_{\hphantom{i}j}^i - \frac{1}{(n-1)} (\operatorname{tr}_g \pi)^2\right)\\
 &\hphantom{=} {} - \mu_g R(g) + 2\Lambda \mu_g = 0,\end{split}\label{eq:117d}
 \end{align}
\end{subequations}
which is equivalent to the system (\ref{eq:106}--\ref{eq:107}). The CMCSH gauge conditions (\ref{eq:109}) now take the form
\begin{subequations}\label{eq:118}
\begin{align}
 \tau &= \operatorname{tr}_g k = \frac{1}{(n-1)} \frac{\operatorname{tr}_g \pi}{\mu_g} = t,\label{eq:118a}\\
 \intertext{and}
 V^k &= g^{ij} \left(\Gamma_{ij}^k(g) - \hat{\Gamma}_{ij}^k(\hat{g})\right) = 0\label{eq:118b}
 \end{align}
\end{subequations}
whereas the elliptic equations for the lapse and shift needed to preserve these gauge conditions become
\begin{subequations}\label{eq:119}
 \begin{align}
 -\Delta_g N + N \left\lbrack\frac{\pi_m^{\operatorname{tr}n} \pi_n^{\operatorname{tr}m}}{(\mu_g)^2} + \frac{\tau^2}{n} - \frac{2\Lambda}{(n-1)}\right\rbrack &= 1,\label{eq:119a}\\
 \begin{split}
 (\Delta_g X)^i + R^i_m(g) X^m - (\mathcal{L}_X V)^i &= \left(\frac{2N}{\mu_g} \left(\pi^{mn} - \frac{1}{(n-1)} (\operatorname{tr}_g \pi) g^{mn}\right) + 2\nabla^m X^n\right)\\
 &\hphantom{=} {} \times \left(\Gamma_{mn}^i(g) - \hat{\Gamma}_{mn}^i(\hat{g})\right)\\
 &\hphantom{=} {} -2(\nabla_m N) \frac{1}{\mu_g} \left(\pi^{im} - \frac{1}{(n-1)} g^{im} (\operatorname{tr}_g \pi)\right)\\
 &\hphantom{=} {} -(\nabla^i N) \left(\frac{\operatorname{tr}_g \pi}{(n-1)\mu_g}\right)\end{split}\label{eq:119b}
 \end{align}
\end{subequations}
where, in the above, we have used the notation
\begin{equation*}
\operatorname{tr}_g \pi := g_{ij} \pi^{ij}
\end{equation*}
for the trace of the momentum \(\pi\) and
\begin{equation}\label{eq:120}
(\pi^{\operatorname{tr}})^{ij} := \pi^{ij} - \frac{1}{n} g^{ij} (\operatorname{tr}_g \pi)
\end{equation}
for its trace free part.

Note that when the CMC gauge condition is enforced one has \(\nabla_i (\operatorname{tr}_g \pi) = 0\) so that the momentum constraint, \(\mathcal{J}_i(g,\pi) = 0\), reduces to
\begin{equation}\label{eq:121}
-2\; \nabla_j(\pi^{\operatorname{tr}})_{\hphantom{j}i}^j = 0.
\end{equation}
Thus \(\pi^{\operatorname{tr}}\) is constrained to be both `transverse' (i.e., divergence free) as well as traceless. Under these circumstances we shall write \(\pi^{\mathrm{TT}}\) for \(\pi^{\operatorname{tr}}\) so that, in particular,
\begin{equation}\label{eq:122}
\pi^{ij} \rightarrow (\pi^{\mathrm{TT}})^{ij} + \frac{1}{n} g^{ij} \operatorname{tr}_g \pi
\end{equation}
and the Hamiltonian constant, \(\mathcal{H}(g,\pi) = 0\), now becomes
\begin{equation}\label{eq:123}
\frac{\pi^{\mathrm{TT}} \cdot \pi^{\mathrm{TT}}}{\mu_g} - \frac{(n-1)}{n} \tau^2 \mu_g + 2\Lambda \mu_g - \mu_g R(g) = 0
\end{equation}
where we have written \(\pi^{\mathrm{TT}} \cdot \pi^{\mathrm{TT}}\) for \((\pi^{\mathrm{TT}})_{\hphantom{n}m}^n (\pi^{\mathrm{TT}})_{\hphantom{m}n}^m\). Note especially that in both of the equations (\ref{eq:119a}) and (\ref{eq:123}) the quantities \(\tau^2\) and \(\Lambda\) appear only in the particular combination \(\tau^2 - 2n\Lambda/(n-1)\). Since we are interested primarily in the case \(\Lambda > 0\) it might appear that this quantity could be negative. But then the Hamiltonian constraint would imply that \(R(g) \geq 0\) everywhere on \textit{M} which is impossible for a manifold of negative Yamabe type. Thus for the spacetimes of interest herein we shall always have
\begin{equation}\label{eq:124}
\tau^2 - \frac{2n\Lambda}{(n-1)} > 0
\end{equation}
and, for expanding universe models,
\begin{equation}\label{eq:125}
-\infty < \tau < -\sqrt{\frac{2n\Lambda}{n-1}}.
\end{equation}

In any dimension \(n \geq 2\) there is a well-known technique, pioneered by Lichnerowicz, for solving the constraint equations on a constant-mean-curvature hypersurface (see Choquet-Bruhat \cite{Choquet-Bruhat:2009} and Bartnik and Isenberg \cite{Bartnik:2004} for detailed expositions of this `conformal' method). If \(n = 2\) and \(M \approx \Sigma_p\) with \(p \geq 2\), or if \(n \geq 3\) and \textit{M} is of negative Yamabe type, then every Riemannian metric \textit{g} on \textit{M} is uniquely globally (pointwise) conformal to a metric \(\gamma\) which satisfies \(R(\gamma) = -1\). In this case every Riemannian metric \textit{g} on \textit{M} can be uniquely expressed as
\begin{equation}\label{eq:126}
g = \begin{cases}
    e^{2\varphi}\gamma & \text{if $n = 2$ and $M \approx \Sigma_p, p \geq 2$}\\
    \varphi^{\frac{4}{(n-2)}}\gamma & \text{if $n \geq 3$ and \textit{M} is of negative Yamabe type}
\end{cases}
\end{equation}
with the conformal metric \(\gamma\) normalized so that \(R(\gamma) = -1\) and where the specific form of the coefficient conformal factor has been chosen to simplify calculations involving curvature tensors. When \(n \geq 3 \; \varphi\) is positive and thus the space of all Riemannian metrics on \textit{M} is parametrized by \(\mathcal{M}_{-1}(M)\) (c.f. Eq.~(\ref{eq:02})) and the  space of scalar functions \(\varphi > 0\) on \textit{M}. 
\section{The Reduced Hamiltonian and its Monotone Decay}
\label{sec:reduced-Hamiltonian-monotone-decay}
Given a scalar function \(\varphi\) (with \(\varphi > 0\) if \(n \geq 3\)) we define, in terms of the `physical' variables \((g, \pi^{\mathrm{TT}})\) a set of `conformal' variables \((\gamma, p^{\mathrm{TT}})\) by setting
\begin{equation}\label{eq:201}
(g, \pi^{\mathrm{TT}}) = \begin{cases}
    (e^{2\varphi}\gamma, e^{-2\varphi} p^{\mathrm{TT}}) & \text{if $n = 2$}\\
    (\varphi^{\frac{4}{(n-2)}}\gamma, \varphi^{-\frac{4}{(n-2)}} p^{\mathrm{TT}}) & \text{if $n \geq 3$}
\end{cases}
\end{equation}
where, as above, \textit{M} is assumed to be of negative Yamabe type, \(R(\gamma) = -1\) and \(\pi^{\mathrm{TT}}\) is transverse-traceless with respect to \textit{g}. It is straightforward to verify that (the symmetric tensor density) \(p^{\mathrm{TT}}\) is transverse-traceless with respect to \(\gamma\), i.e. that
\begin{align}
\gamma_{ij} (p^{\mathrm{TT}})^{ij} &=0, \quad \text{and}\label{eq:202}\\
\nabla_j(\gamma) (p^{\mathrm{TT}})^{ij} &=0, \label{eq:203}
\end{align}
if and only if \(\pi^{\mathrm{TT}}\) is transverse-traceless with respect to \textit{g}. We therefore define a `reduced phase space', \(P_{\mathrm{reduced}}(M)\), of conformal variables by
\(P_{\mathrm{reduced}}(M) = \left\lbrace (\gamma, p^{\mathrm{TT}})| \gamma \in \mathcal{M}_{-1}(M)\right.\) and \(p^{\mathrm{TT}}\) is a 2-contravariant, symmetric tensor density that is transverse and traceless with respect to \(\left.\vphantom{p^{\mathrm{TT}}}\gamma\right\rbrace\).

Any point \((\gamma, p^{\mathrm{TT}}) \in P_{\mathrm{reduced}}(M)\) combined with an \textit{arbitrary} scalar function \(\varphi\) (with \(\varphi > 0\) if \(n \geq 3\)) determines, via (\ref{eq:201}), a solution to the momentum constraint, \(\mathcal{J}_i(g, \pi) = 0\), in CMC gauge (with, for the moment, an \textit{arbitrary constant} choice for the mean curvature \(\tau\))
and indeed every such solution is obtained in this way by allowing \((\gamma, p^{\mathrm{TT}})\) to range over the full reduced phase space \(P_{\mathrm{reduced}}(M)\) and \(\varphi\) to range over the full space of allowed scalar fields on \textit{M}.

The choice of \(\varphi\) however is naturally fixed by now imposing the Hamiltonian constraint, \(\mathcal{H}(g,\pi) = 0\), which, in terms of the conformal variables, takes the form of Lichnerowicz's equation \cite{Choquet-Bruhat:2009,Bartnik:2004}
\begin{equation}\label{eq:204}
\begin{split}
    \mathcal{H}(g, \pi) &= \varphi^{(2n-4)/(n-2)} \mu_\gamma \left(\frac{4(n-1)}{(n-2)} \varphi^{-1} \Delta'_\gamma \varphi - R(\gamma)\right)\\
     & + \varphi^{-2n/(n-2)} \mu_\gamma^{-1} (p^{\mathrm{TT}} \cdot p^{\mathrm{TT}})\\
     & -\varphi^{2n/(n-2)} \mu_\gamma \left(\frac{(n-1)}{n} \tau^2 - 2\Lambda\right) = 0
\end{split}
\end{equation}
for \(n \geq 3\) and
\begin{equation}\label{eq:205}
\begin{split}
    \mathcal{H}(g,\pi) &= \mu_\gamma \left(2\;\Delta'_\gamma \varphi - R(\gamma)\right)\\
    & + e^{-2\varphi} p^{\mathrm{TT}} \cdot p^{\mathrm{TT}} \mu_\gamma^{-1}\\
    & - e^{2\varphi} \mu_\gamma \left(\frac{\tau^2}{2} - 2\Lambda\right) = 0
\end{split}
\end{equation}
for \(n = 2\), where \(\Delta'_\gamma \varphi = \gamma^{ij} \nabla_i(\gamma) \nabla_j(\gamma) \varphi\) and
\begin{equation}\label{eq:206}
\begin{split}
    p^{\mathrm{TT}} \cdot p^{\mathrm{TT}} &= \gamma_{ik} \gamma_{j\ell} (p^{\mathrm{TT}})^{ij} (p^{\mathrm{TT}})^{k\ell}\\
    &= g_{ik} g_{j\ell} (\pi^{\mathrm{TT}})^{ij} (\pi^{\mathrm{TT}})^{k\ell} = \pi^{\mathrm{TT}} \cdot \pi^{\mathrm{TT}}
\end{split}
\end{equation}
Note that in the first term the center dot denotes \(\gamma\)-metric contraction whereas in the last term it denotes \textit{g}-metric contraction. More precisely, we adopt, for convenience, the convention that indices on \(p^{\mathrm{TT}}\) are raised and lowered with respect to \(\gamma\) whereas for \(\pi^{\mathrm{TT}}\) they are raised and lowered with respect to \textit{g}.

By standard methods involving, for example, that of sub- and super-solutions (see, e.g., \cite{Choquet-Bruhat:2009,Bartnik:2004}) one can prove that (\ref{eq:204}) has a unique positive solution \(\varphi = \varphi(\gamma, p^{\mathrm{TT}}, \tau)\) for arbitrary \((\gamma, p^{\mathrm{TT}}) \in P_{\mathrm{reduced}}(M)\) and arbitrary (constant) \(\tau\) satisfying inequality (\ref{eq:124}) and, similarly, that (\ref{eq:205}) has a unique solutions \(\varphi = \varphi (\gamma, p^{\mathrm{TT}}, \tau)\) for arbitrary \((\gamma, p^{\mathrm{TT}})\) and arbitrary constant \(\tau\) such that \(\tau^2 - 4\Lambda > 0\). By contrast no solution exists if these inequalities are violated since then Eqs.~(\ref{eq:204}--\ref{eq:205}) would imply that \(\Delta'_\gamma \varphi < 0\) on \textit{M} which is impossible since \textit{M} is compact. To obtain solutions with \(\tau^2 -2\Lambda/(n-1) \leq 0\) would require that one relax the condition that \textit{M} be of negative Yamabe type.

We focus on negative values of \(\tau\) (c.f., Eq.~(\ref{eq:125})) since, with our convention, these correspond to expanding universes whereas positive values for \(\tau\) would correspond to time-reversed, collapsing universes. To derive the reduced Hamiltonian we begin with the ADM (Arnowitt, Deser and Misner \cite{Arnowitt:1962,Misner:1973}) action functional \(I_{\mathrm{ADM}}\) defined on the cylinder \(\bar{I} \times M\), where \(\bar{I} = [t_o,t_1] \subset R\), by
\begin{equation}\label{eq:207}
I_{\mathrm{ADM}} = \int_{\bar{I}} dt \int_M d^nx \left\lbrace\pi^{ij} \partial_t g_{ij} - N\mathcal{H}(g,\pi) - X^i \mathcal{J}_i(g,\pi)\right\rbrace.
\end{equation}
Substituting an arbitrary differentiable curve of solutions to the constraints into this expression after imposing the CMC condition that \(\tau = \tau(t)\) one gets, upon noting that
\begin{equation}\label{eq:208}
\pi^{ij} \partial_t g_{ij} = (p^{\mathrm{TT}})^{ij} \partial_t \gamma_{ij} - \frac{2(n-1)}{n} (\partial_t\tau) \mu_g + \frac{2}{n} \partial_t (\operatorname{tr}_g \pi),
\end{equation}
the following expression for the reduced action
\begin{equation}\label{eq:209}
I_{\mathrm{reduced}} = \int_{\bar{I}} dt \int_{M} d^n x \left((p^{\mathrm{TT}})^{ij} \partial_t\gamma_{ij} - \frac{2(n-1)}{n} (\partial_t\tau) \mu_g\right)
\end{equation}
where we have discarded a boundary term of the form \(\int_M \frac{2}{n} (\operatorname{tr}_g \pi) d^nx |_{t_0}^{t_1}\) since this will not contribute to the resulting equations of motion.

We now choose a time coordinate \textit{t} so that
\begin{equation}\label{eq:210}
\frac{d\tau}{dt} = \frac{n}{2(n-1)} \left(\tau^2 - \frac{2n\Lambda}{(n-1)}\right)^{n/2}
\end{equation}
and the reduced action becomes
\begin{equation}
I_{\mathrm{reduced}} = \int_{\bar{I}} dt \int_M d^n x \left\lbrack (p^{\mathrm{TT}})^{ij} \partial_t\gamma_{ij} - \left(\tau^2 - \frac{2n\Lambda}{(n-1)}\right)^{n/2} \mu_g\right\rbrack
\end{equation}
from which we can read off the effective \textit{reduced Hamiltonian}
\begin{equation}\label{eq:211}
\begin{split}
    H_{\mathrm{reduced}} &= \int_M \left(\tau^2 - \frac{2n\Lambda}{(n-1)}\right)^{n/2} d\mu_g\\
    &= \int_M \left(\tau^2 - \frac{2n\Lambda}{(n-1)}\right)^{n/2} \varphi^{\frac{2n}{(n-2)}} d\mu_\gamma
\end{split}
\end{equation}
for \(n \geq 3\) and
\begin{equation}\label{eq:215}
H_{\mathrm{reduced}} = \int_{\Sigma_p} (\tau^2 - 4\Lambda) e^{2\varphi} d\mu_\gamma
\end{equation}
if \(n = 2\). Here \(\varphi = \varphi (\tau, \gamma, p^{\mathrm{TT}})\) is that positive functional of the reduced phase space variables determined uniquely by the corresponding Lichnerowicz equation ((\ref{eq:204}) or (\ref{eq:205}) respectively).

Note that, strictly speaking, the reduced action is not in canonical Hamiltonian form until we restrict \(\gamma\) further to lie in a local \(\mathcal{D}_0\)-cross-section of \(\mathcal{M}_{-1}\) which represents a local chart for the quotient space \(\mathcal{M}_{-1}(M)/\mathcal{D}_0(M)\). Such a cross-section has, at any point \(\gamma\), a tangent space modeled on the space of transverse-traceless tensors relative to \(\gamma\) and thus a `dimension' which matches that of the cotangent bundle fiber space whose elements are of the momenta \(p^{\mathrm{TT}}\). Thus in the fully reduced setting the isometry class \(\lbrack\gamma\rbrack\) ranges over the orbit space \(\mathcal{M}_{-1}(M)/\mathcal{D}_0(M)\) and the pair \(\left(\lbrack\gamma\rbrack, p^{\mathrm{TT}}\right) \in P_{\mathrm{reduced}}\) represents `coordinates' for \(T^\ast \left(\mathcal{M}_{-1}(M)/\mathcal{D}_0(M)\right)\). This additional restriction is not needed in most of what follows since we are mainly interested in computing the behavior of \(\mathcal{D}_0\)-invariant quantities such as the  reduced Hamiltonian itself and for that reason we shall refrain from enforcing it to simplify the analysis.

The reduced Hamiltonian depends explicitly upon the time variable \textit{t} both through the multiplicative factor \(\left(\tau^2(t) - 2n\Lambda/(n-1)\right)^{n/2}\) and through the fact that the conformal factor depends upon \(\tau(t)\) in view of the latter's explicit occurrence in the Lichnerowicz equation. Thus the actual reduced dynamics takes place on the corresponding contact manifold \(\approx R \times T^\ast \left(\mathcal{M}_{-1}(M)/\mathcal{D}_0(M)\right)\) and the reduced Hamiltonian is not a conserved quantity. In fact, as we shall see below, \(H_{\mathrm{reduced}}\) is universally monotonically decaying except on very special solutions which exist only if \textit{M} admits a negative Einstein metric in which case \(H_{\mathrm{reduced}}\) is constant.

Again, since we are dealing primarily with \(\mathcal{D}_0\)-invariant quantities, we shall take
\begin{equation}\label{eq:216}
\bar{I} \times P_{\mathrm{reduced}} = \left\lbrace(\tau, \gamma, p^{\mathrm{TT}})|\tau \in \bar{I} = \left(-\infty, -\sqrt{\frac{2n\Lambda}{(n-1)}}\right), (\gamma , p^{\mathrm{TT}}) \in P_{\mathrm{reduced}}\right\rbrace
\end{equation}
as the `contact' manifold on which \(H_{\mathrm{reduced}}\) is defined.

A straightforward calculation, using the ADM form of the Einstein-\(\Lambda\) field eqs.~(c.f.~\ref{eq:117}) specialized to CMC gauge and the time coordinate \textit{t} defined via (\ref{eq:210}), yields
\begin{equation}\label{eq:217}
\frac{dH_{\mathrm{reduced}}}{dt} = \int_M N\tau \left(\tau^2 - \frac{2n\Lambda}{(n-1)}\right)^{(n/2)-1} n\left(\frac{\pi^{\mathrm{TT}} \cdot \pi^{\mathrm{TT}}}{(\mu_g)^2}\right)d\mu_g
\end{equation}
where we have exploited the elliptic equation for the lapse function \textit{N} given by
\begin{equation}\label{eq:218}
\begin{split}
    \frac{\partial\tau}{dt} &= \frac{n}{2(n-1)} \left(\tau^2 - \frac{2n\Lambda}{(n-1)}\right)^{n/2}\\
     &= -\Delta_g N + N \left\lbrack\frac{\pi^{\mathrm{TT}} \cdot \pi^{\mathrm{TT}}}{(\mu_g)^2} + \frac{\tau^2}{n} - \frac{2\Lambda}{(n-1)}\right\rbrack
\end{split}
\end{equation}
that is needed to preserve the CMC slicing condition. A standard maximum principle argument applied to (\ref{eq:218}) shows that \(N > 0\) on \textit{M} whereas since \(\pi^{\mathrm{TT}} \cdot \pi^{\mathrm{TT}} = (\pi^{\mathrm{TT}})_m^{\hphantom{m}n}(\pi^{\mathrm{TT}})_n^n \geq 0\) and \(\tau < 0\) we clearly get
\begin{equation}\label{eq:219}
\frac{dH_{\mathrm{reduced}}}{dt} \leq 0
\end{equation}
with equality holding (perhaps only instantaneously) if and only if \(\pi^{\mathrm{TT}} = 0\) at that instant.

We now go further to show that \(H_{\mathrm{reduced}}\) is actually strictly monotonically decreasing except for a set of extremely special solutions for which the conformal metric \(\gamma\) is a fixed (i.e, time-independent) Einstein metric and \(H_{\mathrm{reduced}}\) is constant.

Suppose that at some instant \(\pi^{\mathrm{TT}}\) did vanish. Then clearly from (\ref{eq:217}), both \(\frac{dH_{\mathrm{reduced}}}{dt}\) and \(\frac{d^2H_{\mathrm{reduced}}}{dt^2}\) vanish at that instant, whereas
\begin{equation}\label{eq:220}
\frac{d^3H_{\mathrm{reduced}}}{dt^3} = \int_M N\tau \left(\tau^2 - \frac{2n\Lambda}{(n-1)}\right)^{(n/2)-1} n\; \frac{\partial_t\pi^{\mathrm{TT}} \cdot \partial_t\pi^{\mathrm{TT}}}{(\mu_g)^2} d\mu_g
\end{equation}
which is strictly negative unless \(\partial_t\pi^{\mathrm{TT}} = 0\) also at that instant.

Thus suppose that on some CMC slice we have \(\pi^{\mathrm{TT}} = 0\) and \(\partial_t\pi^{\mathrm{TT}} = 0\) simultaneously. The elliptic equation (\ref{eq:218}) for the lapse then has the unique solution
\begin{equation}\label{eq:221}
N = \frac{n^2}{2(n-1)} \left(\tau^2 - \frac{2n\Lambda}{(n-1)}\right)^{\frac{n}{2}-1}
\end{equation}
and the ADM field equations (\ref{eq:117}) can be combined to yield
\begin{equation}\label{eq:222}
R_{ij}(g) = -\frac{(n-1)}{n^2} \left(\tau^2 - \frac{2n\Lambda}{(n-1)}\right) g_{ij}
\end{equation}
so that \textit{g} is necessarily a negative Einstein metric on the chosen slice. When \(\pi^{\mathrm{TT}} = 0\) the unique, positive solution to the Lichnerowicz equation is easily shown to be given by
\begin{equation}\label{eq:223}
\varphi^{4/(n-2)} = \frac{-R(\gamma)}{\left(\frac{(n-1)}{n} \tau^2 - 2\Lambda\right)} = \frac{1}{\left(\frac{(n-1)}{n} \tau^2 - 2\Lambda\right)}
\end{equation}
for \(n \geq 3\), and
\begin{equation}\label{eq:224}
e^{2\varphi} = \frac{-R(\gamma)}{\left(\frac{\tau^2}{2} - 2\Lambda\right)} = \frac{1}{\left(\frac{\tau^2}{2} - 2\Lambda\right)}
\end{equation}
when \(n = 2\) so that \(\gamma = \varphi^{-4/(n-2)}g\) or \(\gamma = e^{-2\varphi}g\), respectively, satisfies
\begin{equation}\label{eq:225}
R_{ij}(\gamma) = -\frac{1}{n}\gamma_{ij}\qquad \forall\; n \geq 2.
\end{equation}

When \(\pi^{\mathrm{TT}} = 0, \frac{\operatorname{tr}_g\pi}{\mu_g} = (n-1)\tau = \hbox{constant}\) and \textit{g} is an Einstein metric satisfying (\ref{eq:222}) on an initial CMC Cauchy hypersurface the full set of field equations may be integrated explicitly to yield the warped-product solution given by
\begin{equation}\label{eq:226}
ds^2 = \frac{-n^2}{\left(\tau^2 - \frac{2n\Lambda}{(n-1)}\right)^2} d\tau^2 + \frac{1}{\left(\frac{(n-1)}{n} \tau^2 - 2\Lambda\right)} \gamma_{ij} dx^i dx^j
\end{equation}
where \(\gamma = \gamma_{ij} dx^i \otimes dx^j\) is a fixed (i.e., \(\tau\)-independent) Einstein metric on \textit{M} satisfying (\ref{eq:225}) so that, in particular, \(R(\gamma) = -1\).

The reduced Hamiltonian is easily verified to be constant (i.e., `time'-independent) when evaluated on these `warped-product' solutions (\ref{eq:226}) which thus provide the only Einstein-\(\Lambda\) spacetimes on which it fails to be strictly monotonically decaying in the temporal direction of cosmological expansion. Each of these special solutions admits a (future directed) timelike, conformal Killing field \textit{Y} given by
\begin{equation}\label{eq:227}
Y = Y^\alpha \frac{\partial}{\partial x^\alpha} = \left(\tau^2 - \frac{2n\Lambda}{(n-1)}\right)^{1/2} \frac{\partial}{\partial\tau}
\end{equation}
with
\begin{equation}\label{eq:228}
\mathcal{L}_Y \bar{g} = \frac{-2\tau}{\left(\tau^2 - \frac{2n\Lambda}{(n-1)}\right)^{1/2}} \bar{g}
\end{equation}
where \(\bar{g} = \bar{g}_{\mu\nu} dx^\mu \otimes dx^\nu\) is a metric with line element given by (\ref{eq:101}). In the special case for which \(\Lambda = 0\) \textit{Y} reduces to the \textit{homothetic} Killing field \(Y = -\tau \frac{\partial}{\partial\tau}\) with \(\mathcal{L}_Y\bar{g} = 2\bar{g}\) and the associated spacetimes are the self-similar `Lorentz cone'  ones analyzed in Refs.~\cite{Fischer:2002,Fischer:2002b,Andersson:2011}.

When \(\Lambda \neq 0\) one can show, by methods virtually identical to those of Section 4.2 in \cite{Fischer:2002b},\footnote{One need only make the replacement of \(\tau^2\) in the \(\Lambda = 0\) argument with \(\tau'^2 := \tau^2 - \frac{2n\Lambda}{(n-1)}\) to handle the cases for which \(\Lambda > 0\).\label{note02}} that the conformal data points \((\gamma, p^{\mathrm{TT}}) \in P_{\mathrm{reduced}}\) such that \(p^{\mathrm{TT}} = 0\) and \(\gamma\) is an Einstein metric with \(R(\gamma) = -1\) are precisely the \textit{critical points} of the reduced Hamiltonian and thus precisely the fixed points of the corresponding Hamiltonian flow.

In the special case \(n = 2\), \(M = \Sigma_p\), \(p \geq 2\), a simple formula for \(H_{\mathrm{reduced}}\) can be derived. By combining (\ref{eq:126}), (\ref{eq:205}) and (\ref{eq:215}) we find that
\begin{equation}\label{eq:229}
\begin{split}
    H_{\mathrm{reduced}} (\tau, \gamma, p^{\mathrm{TT}}) &= \int_{\Sigma_p} (\tau^2 - 4\Lambda) e^{2\varphi(\tau, \gamma, p^{\mathrm{TT}})} d\mu_\gamma\\
    &= \int_{\Sigma_p} (\tau^2 - 4\Lambda) d\mu_g\\
    &= 2\int_{\Sigma_p} (\det{(e^{2\varphi}\gamma)})^{-1} p^{\mathrm{TT}} \cdot p^{\mathrm{TT}} d\mu_{(e^{2\varphi}\gamma)} - 2\int_{\Sigma_p} R(g) d\mu_g\\
    &= 2\int_{\Sigma_p} e^{-2\varphi} (\det{\gamma})^{-1} (p^{\mathrm{TT}} \cdot p^{\mathrm{TT}}) d\mu_\gamma - 8\pi\; \chi (\Sigma_p)\\
    &= 2\int_{\Sigma_p} e^{-2\varphi} (\det{\gamma})^{-1} (p^{\mathrm{TT}} \cdot p^{\mathrm{TT}}) d\mu_\gamma + 16\pi\; (p-1)
\end{split}
\end{equation}
where \(\varphi = \varphi (\tau, \gamma, p^{\mathrm{TT}})\) is the solution to the Lichnerowicz equation (\ref{eq:205}), \(\chi (\Sigma_p) = 2(1 - p)\) is the Euler characteristic of the genus \textit{p} surface \(\Sigma_p\), and where we have used the Gauss-Bonnet theorem
\begin{equation}\label{eq:230}
\int_{\Sigma_p} R(g) d\mu_g = 4\pi\;\chi (\Sigma_p) = 8\pi\; (1 - p)
\end{equation}
to simplify the resulting formula.

Since
\begin{equation}\label{eq:231}
H_{\mathrm{reduced}} (\tau, \gamma, p^{\mathrm{TT}}) = 2\int_{\Sigma_p} e^{-2\varphi} (\det{\gamma})^{-1} (p^{\mathrm{TT}} \cdot p^{\mathrm{TT}}) d\mu_\gamma + 16\pi\; (p - 1) \geq 16\pi \;(p - 1)
\end{equation}
the infimum of \(H_{\mathrm{reduced}}\) is attained precisely when \(p^{\mathrm{TT}} = 0\) and this infimum coincides with the topological invariant \(-8\pi\;\chi (\Sigma_p) = 16\pi \;(p - 1)\) which characterizes the surface \(\Sigma_p\). As we shall see shortly an analogous result holds for \(n \geq 3\).

Using special methods, applicable only for \(n = 2\), one can prove that \textit{every solution} to the Einstein-\(\Lambda\) field equations (for \(\Lambda \geq 0\)) evolves so that the infimum of \(H_{\mathrm{reduced}}\) is asymptotically achieved in the limit as \(\tau \nearrow 2\sqrt{\Lambda}\) (i.e., in the limit of infinite `volume' expansion) \cite{Moncrief:2008,Andersson:1997}. While no such general `global existence' result is available in the higher dimensional cases one can nevertheless rigorously analyze the (fully-nonlinear) stability of those special solutions (\ref{eq:226}) that exist when \textit{M} admits a negative Einstein metric \(\gamma\) \cite{Andersson:2011,Andersson:2004}. Indeed, for the special cases with \(\Lambda = 0\) one can prove that these \textit{n}-dimensional Einstein spaces \(\lbrace M, \gamma\rbrace\) are, in a natural dynamical sense, `attractors' for the (vacuum) Einstein `flow' \cite{Andersson:2011} and a result of the same type is anticipated to hold when \(\Lambda > 0\) as well.

Here however we are more interested in those `exotic' topologies in higher dimensions for which no Einstein metric exists at all. This is therefore the topic to which we now return. 
\section{The Infimum of $H_{\mathrm{reduced}}$ and the $\sigma$-constant of \textit{M}}
\label{sec:infimum}
Since the reduced Hamiltonian is bounded from below (as a rescaled volume of CMC hpersurfaces) and is universally monotonically decaying in the direction of cosmological expansion (except for the `warped product' solutions (\ref{eq:226}) on which it is constant) it is natural to ask what its infimum is and whether this infimum is ever attained, at least asymptotically, by solutions of the field equations. The answer, in part, is provided by a theorem that characterizes the infimum of \(H_{\mathrm{reduced}}\) (taken, at fixed \(\tau\), over all of \(T^\ast \mathfrak{T}(M)\)) in terms of a topological invariant known as the \(\sigma\)-constant (or Yamabe invariant) of \textit{M} \cite{Schoen:1989,Anderson:1997}:
\begin{equation}\label{eq:301}
\inf_{(\gamma, p^{\mathrm{TT}}) \in T^\ast\mathfrak{T}(M)} H_{\mathrm{reduced}} (\tau, \gamma, p^{\mathrm{TT}}) = \left\lbrack\left(\frac{n}{(n-1)}\right) \left(-\sigma (M)\right)\right\rbrack^{n/2}.
\end{equation}
The \(\sigma\) constant is, in a sense, a natural generalization of the Euler characteristic \(\chi (\Sigma_p)\) of a compact surface since, when restricted to two dimensions, its definition leads to:
\begin{equation}\label{eq:302}
\sigma (\Sigma_p) = 4\pi\; \chi (\Sigma_p) = 8\pi\; (1 - p).
\end{equation}
More generally, for manifolds of negative Yamabe type in higher dimensions, the precise definition leads to the formula:
\begin{equation}\label{eq:303}
\sigma (M) = -\left(\inf_{\gamma \in \mathcal{M}_{-1}(M)} \operatorname{vol} (M, \gamma)\right)^{2/n}
\end{equation}
where \(\operatorname{vol} (M, \gamma) := \int_M d\mu_\gamma\) (c.f., Section 4.6 of Ref.~\cite{Fischer:2002b} for a more extensive discussion of this formula).

To compute the infimum of
\begin{equation}\label{eq:304}
H_{\mathrm{reduced}} (\tau, \gamma, p^{\mathrm{TT}}) = \int_M \left(\tau^2 - \frac{2n\Lambda}{(n-1)}\right)^{n/2} \varphi^{2n/(n-2)} (\tau, \gamma, p^{\mathrm{TT}}) d\mu_\gamma
\end{equation}
for \(n \geq 3\) we first fix \(\tau \in \left(-\infty, -\sqrt{\frac{2n\Lambda}{(n-1)}}\right)\) and \(\gamma \in \mathcal{M}_{-1}(M)\) and vary the fibre variable \(p^{\mathrm{TT}}\) of the cotangent bundle
\begin{equation}\label{eq:305}
T^\ast \left(\mathcal{M}_{-1}(M)/\mathcal{D}_0(M)\right) = T^\ast\mathfrak{T}(M).
\end{equation}
A straightforward maximum principle argument, applied to the Lichnerowicz equation (\ref{eq:204}), shows that the unique positive solution \(\varphi = \varphi (\tau, \gamma, p^{\mathrm{TT}})\) satisfies
\begin{equation}\label{eq:306}
\varphi^{2n/(n-2)} \geq \left(\frac{n}{(n-1)}\right)^{n/2} \frac{1}{\left(\tau^2 - \frac{2n\Lambda}{(n-1)}\right)^{n/2}}
\end{equation}
with equality holding everywhere on \textit{M} if and only if \(p^{\mathrm{TT}}\) vanishes identically on \textit{M}. In that case \(H_{\mathrm{reduced}}\) simplifies to
\begin{equation}\label{eq:307}
H_{\mathrm{reduced}} (\tau, \gamma, 0) = \left(\frac{n}{n-1}\right)^{n/2} \int_M d\mu_\gamma
\end{equation}
which thus is now independent of \(\tau\) and depends only on the volume of \(\gamma \in \mathcal{M}_{-1}(M)\). It follows that, for arbitrary but fixed \(\tau \in \left(-\infty, - \sqrt{\frac{2n\Lambda}{(n-1)}}\right)\),
\begin{equation}\label{eq:308}
\begin{split}
    \inf_{T^\ast\mathfrak{T}(M)} H_{\mathrm{reduced}} &= \left(\frac{n}{n-1}\right)^{n/2} \inf_{\gamma \in \mathcal{M}_{-1}(M)} \int_M d\mu_\gamma\\
    &= \left(\frac{n}{n-1}\right)^{n/2} \left(-\sigma (M)\right)^{n/2}
\end{split}
\end{equation}
thereby confirming the statement (\ref{eq:301}) made above.

If matter sources obeying a suitable energy condition are allowed the argument goes through in much the same way as above in that the \textit{rescaled volume} (which need however no longer be an actual Hamiltonian for the augmented field equations) is still monotonically decaying in the direction of cosmological expansion and has the aforementioned infimum only in the limit that the matter sources be `turned off' or, at least, become asymptotically negligible.

It has long been realized that a \textit{graph} 3-manifold G has $\sigma (G) = 0$ since, roughly speaking, a sequence of conformal metrics seeking to achieve the indicated infimum tends to collapse its circular or $\Sigma_p$  fibers. Thus no actual metric on G has a volume that realizes the $\sigma$ constant; the latter can only be approached in a degenerating limit. Thanks to the recent progress in Ricci flow, however, it is now known that the  $\sigma$ constant of a \textit{hyperbolizable} manifold is actually achieved by its hyperbolic metric. Using different methods some of the  $\sigma$ constants of positive Yamabe type manifolds have also been computed \cite{Bray:2004}.

Of most interest to us however is the fact that Ricci flow techniques have been used to determine the $\sigma$ constant (and therefore the infimum of the reduced Hamiltonian) of the most general compact 3-manifold of \textit{negative Yamabe type}. The result is given simply by:
\begin{equation}
|\sigma (M)| = (\mathrm{vol}_{-1} H)^{2/3}
\label{eq:09}
\end{equation}
where $\mathrm{vol}_{-1}\; H$  is the volume of the hyperbolic part of M computed with respect to the hyperbolic metric normalized to have scalar curvature $=  -1$  \cite{Anderson:2004,Anderson:2006}. In particular, any graph manifolds G, spherical space forms $S^3 /\Gamma_i$   or handles  $S^2 \times S^1$, even if present in M, make no contribution to the sigma constant of M and hence none as well to the infimum of the reduced Hamiltonian.

Since the reduced Hamiltonian, which geometrically is nothing but the rescaled spatial
volume of the expanding universe model, is universally monotonically decaying towards its
infimum and since that infimum is determined entirely by the hyperbolic component or
components of M we are naturally led to the conclusion sketched in the introduction that
Einstein's equations potentially incorporate a dynamical mechanism for driving the universe
model to an asymptotic state that is volume dominated by hyperbolic components equipped
with their canonical, locally homogeneous and isotropic metrics.

\section{Stability Results --- The Vacuum Limit}
\label{sec:stability}

	To decide the extent to which the reduced Hamiltonian actually does decay to its infimum (or is instead perhaps obstructed from doing so) is a very demanding open problem on the global properties of solutions to the field equations. Aside from some highly symmetric examples (e.g., vacuum Bianchi models) for which one can do explicit calculations \cite{Fischer:2000,Fischer:2006} or in $2 + 1$ dimensions, wherein one can verify the expected, decay-to-infimum behavior through the use of special techniques \cite{Moncrief:2008}, available results are currently limited to stability theorems for  some rather special families of 'background' solutions and to theorems which assume a priori bounds upon spacetime curvature \cite{Anderson:2001,Anderson:2004b,Reiris:2010,Reiris:2009}. An important class of solutions for which dynamical stability results can be proven directly is provided by the vacuum, self-similar `Lorentz cone' spacetimes discussed above following Eq.~(\ref{eq:228}).

	Einstein's vacuum field equations, written in their conventional form, are an autonomous system of partial differential equations for the spacetime metric. When the gauge is fixed by the CMC slicing condition however this autonomous character is apparently broken since both the constraint and evolution equations, as well as the associated elliptic equation for the lapse function, all depend explicitly upon the mean curvature, which is now playing the role of `time'. For the vacuum equations (or in the presence of scale invariant matter sources but not including a cosmological constraint) however one can restore the autonomous character of the gauge-fixed field equations by rewriting them in terms of suitable rescaled, dimensionless variables, using appropriate powers of the mean curvature as scale factors \cite{Fischer:2002,Fischer:2002b,Andersson:2011}. The natural, dimensionless time coordinate for the rescaled equations is now given by $T = -\ln ( \tau /  \tau_0 )$ and has maximal range $(-\infty, \infty)$ and thus serves as an effective `Newtonian time' for this reduced, newly autonomous system.\footnote{Though the reduced system is autonomous the rescaled variables are not strictly canonical so there is no reason to expect the corresponding Hamiltonian to be conserved.\label{note03}}

	When this reformulation is carried out on a \textit{hyperbolic} 3-manifold M (or one admitting a negative Einstein metric in higher dimensions) the resulting dynamical system has the Lorentz cone solutions described previously as its unique fixed points \cite{Fischer:2002b,Andersson:2011}. Since, moreover, in 3 + 1 dimensions these solutions are known to realize the infimum of the reduced Hamiltonian it is natural to ask whether these isolated fixed points (in the reduced phase space) are in fact actual attractors for the associated, reduced Einstein flow. If so then at least sufficiently nearby solutions (in a suitable function space setting) will indeed tend asymptotically to approach the same infimum for $H_{\mathrm{reduced}}$ and, more significantly, the rescaled spatial metric will tend to approach the (locally homogeneous and isotropic) hyperbolic one in the limit of infinite cosmological expansion.

	As a first step towards establishing this conclusion one can analyze the linearized field equations, taking an arbitrary Lorentz cone solution as the background to perturb. While the results of such analyses confirm one's expectations \cite{Fischer:2002b,Andersson:2011} they fall mathematically short of proving the conjectured property for the full nonlinear Einstein flow. For that purpose one needs to develop more sophisticated techniques. The vacuum field equations in $2 + 1$ dimensions are so special (primarily in excluding the possibility of gravitational waves) that one can actually resolve this conjecture (affirmatively) for arbitrarily large perturbations away from the self-similar, Lorentz cone `backgrounds' \cite{Moncrief:2008,Andersson:1997,Benedetti:2001,Andersson:2005}. In $3 + 1$ and higher dimensions however the currently available methods of stability analysis require a certain smallness condition on the nonlinear perturbations for their successful implementation. These methods proceed by defining suitable `energy' functionals that, while positive for nontrivial perturbations actually vanish on the backgrounds and bound the norms needed for control of the existence times of `nearby', perturbed solutions. One aims to show that the appropriate energy functional decays asymptotically to zero, in the direction of cosmological expansion, for any solution whose `initial values' (at some nonzero value $\tau_0 < 0$ of the mean curvature) are sufficiently close to the those of the background and to deduce therefrom the desired stability result.

	However even the local (i.e. short time) existence of solutions in CMC slicing is not covered by the classical existence and uniqueness theorem for Einstein's equations \cite{Foures-Bruhat:1952} since this theorem assumes a \textit{spacetime harmonic} (or \textit{Lorentz} type) gauge condition to reduce the field equations to hyperbolic form. By imposing instead only the aforementioned \textit{spatial harmonic} (or \textit{Coulomb} type) gauge condition to supplement the CMC time-slicing condition one arrives at an elliptic-hyperbolic system of field equations for which, however, a well-posedness theorem for the vacuum (i.e., \(\Lambda = 0\)) equations was established in \cite{Andersson:2003}. In $n + 1$ dimensions, for $n > 2$, this theorem requires (as does the traditional one) the metric to lie in a Sobolev space for which $s > n/2 + 1$ of its (spatial) derivatives are square integrable over M. To extend this local existence result to a global one one needs to prove that the corresponding Sobolev norm of a solution cannot blowup in a finite time. This will be possible whenever one can make the energy arguments alluded to above work in practice.

	One rather geometrically elegant implementation of this program involves defining certain, higher-order \textit{Bel-Robinson} type energy functionals that consist essentially of Sobolev norms of spacetime curvature. These can be employed to verify the anticipated dynamical stability for all hyperbolic 3-manifolds except the (nonempty, proper) subset admitting so-called nontrivial traceless Codazzi tensors \cite{Andersson:2004}. Any member of this latter subset allows a certain finite dimensional moduli space of nontrivial but still flat spacetime perturbations (that are not however of self-similar type). These are invisible to the curvature based Bel-Robinson  energies and so cannot be controlled by them. One can either fill this gap by a separate independent argument or instead develop non-curvature-based energies to handle the full range of possibilities more uniformly \cite{Andersson:2011}.

	This latter approach can be made to work as well in higher dimensions when the background, self-similar solution is a Lorentz cone over an arbitrary (negative) Einstein metric (that need no longer be hyperbolic) provided that the spectrum of its associated elliptic, Lichnerowicz Laplacian satisfies a suitable condition \cite{Andersson:2011}. In this more general setting a finite dimensional space of Einstein metrics provides the `center manifold' towards which the rescaled spatial metric is flowing in the limit of infinite cosmological expansion. All of the  spacetimes that can be handled in this way (as sufficiently small perturbations of self-similar backgrounds that satisfy the needed spectral condition) can be shown to be causally geodesically complete in this same temporal direction. Large families of such backgrounds (and their perturbations) can be constructed by taking Riemannian products of negative Einstein spaces that satisfy the needed spectral condition and verifying that the spectral condition is automatically preserved in the process \cite{Andersson:2011}.

	Energy arguments of the same general type as those described above had, even earlier, been shown to be applicable to U(1)-symmetric vacuum  metrics defined on circle bundles over higher genus surfaces \cite{Choquet-Bruhat:2001,Choquet-Bruhat:2001b,Choquet-Bruhat:2004}. Though limited at the outset to spacetimes having a spacelike Killing symmetry (generating the assumed U(1) action) these results are especially intriguing in the challenge they provoke for an attack on the corresponding \textit{large data} stability problem. Large data global existence results are currently available (in the vacuum, cosmological setting under discussion here) only for so-called Gowdy spacetimes which, by definition, have (spacelike) $U(1) \times U(1)$ isometry groups \cite{Ringstrom:2010} or spacetimes (such as Bianchi models) that have even higher symmetry \cite{asymptotic}. Genuine progress on the actual, large data U(1) problem would  represent a `quantum leap' forward in one's understanding of such issues and therefore deserves a major effort.

	In it's basic form the vacuum U(1) problem can be expressed (through a variant of Kaluza-Klein reduction) as the $2 + 1$ dimensional Einstein equations coupled to a wave map with (two-dimensional) hyperbolic target. The global existence problem for such wave maps on a fixed ($2 + 1$ dimensional) Minkowski background has recently been solved \cite{Schlag:2012,Tataru:2010}. In the simplest, so-called `polarized' case however, which requires that the bundle be trivial for its formulation, the wave map  reduces to a wave equation. The global existence of such (linear) wave equations is of course already well-established even on curved (globally hyperbolic) backgrounds \cite{Friedlander:1975}. To handle the fully coupled U(1)-symmetric field equations though requires simultaneous control over the wave map (or wave equation) and the Teichm\"{u}ller parameters of the $2 + 1$ dimensional Lorentz metric which now is no longer a given background. While it is not currently known how to do this it seems encouraging that the formation of black holes in such spacetimes is obstructed by the imposed symmetry. It thus seems plausible to conjecture that every solution should exist for the maximum possible range of its geometrically defined (CMC) time and, in particular, to expand forever without developing singularities to the future.

	It does not seem likely however that such large data global existence questions can be settled (either for the U(1) problem or, a fortiori, for the fully general non-symmetric one) by pure (higher order) energy arguments. The reduced Hamiltonian is always at hand, and applicable to arbitrarily large data, but can only bound, in principle, an $H^1(M) \times L^2 (M)$ Sobolev-type norm of the reduced phase space variables $\{\gamma,  p^{TT}\}$. The best available local existence theorem (for the general, non-symmetric problem), on the other hand, requires these variables to lie in the higher order $H^2(M) \times H^1 (M)$, or Bel-Robinson energy level, Sobolev space \cite{Klainerman:unpub,Wang:unpub}. But the Bel-Robinson energy, unlike the monotonically decaying reduced Hamiltonian, is \textit{not}, a priori, under control.

	There is however a rather ambitious program under development to control not only the  Bel-Robinson energy but also the pointwise (or  $L^\infty$ -- norm) behavior of spacetime curvature through the use of what we shall informally refer to as \textit{light-cone estimates}. We shall briefly outline one particular variant of this far-reaching program in the final section below.

\section{An Integral Equation for Spacetime Curvature}
\label{sec:integral-spacetime}

	It has long been realized that the Yang-Mills equations, especially when formulated in a curved background spacetime, have many similarities to the Einstein equations and thus, since methods are already at hand for bounding Yang-Mills curvature \cite{Eardley:1982,Eardley:1982b,Crusciel:1997,Klainerman:2007}, similar techniques might well be applicable to the Einstein problem. These similarities are most pronounced when Einstein's theory is expressed in the Cartan, orthonormal frame formalism wherein the Riemann curvature tensor appears as a matrix of two-forms $\{R^{\hat{a}}_{\hphantom{\hat{a}}\hat{c}\mu\nu} dx^\mu \wedge dx^\nu\}$ expressible in terms of the matrix of (Lorentz connection) one-forms  $\{\omega^{\hat{a}}_{\hphantom{\hat{a}}\hat{c}\mu} dx^\mu\}$ via
\begin{equation}\label{eq:10}
\begin{split}
& R^{\hat{c}}\,_{\hat{a}\mu\nu} = \theta^{\hat{c}}_\gamma
h^\lambda_{\hat{a}} R^\gamma \,\,_{\lambda \mu\nu} \\
& = \partial_\mu \omega^{\hat{c}}\,\,_{\hat{a} \nu} - \partial_\nu
\omega^{\hat{c}}\,\,_{\hat{a} \mu} + \omega^{\hat{c}}\,\, _{\hat{d}
\mu} \omega^{\hat{d}}\,\,_{\hat{a} \nu} - \omega^{\hat{c}}\,\,_{
\hat{d} \nu} \omega^{\hat{d}}\,\,_{\hat{a} \mu}.
\end{split}
\end{equation}
Here $h_{\hat{a}} = h_{\hat{a}}^\mu \frac{\partial}{\partial x^\mu}$ and $\theta^{\hat{a}} = \theta^{\hat{a}}_\mu dx^\mu $ are the orthonormal frame and co-frame fields which determine the Lorentz connection by means of the vanishing torsion condition
\begin{equation}
\partial_\nu\theta^{\hat{c}}_\mu - \partial_\mu \theta^{\hat{c}}_\nu +
\omega^{\hat{c}}\,_{\hat{a}\nu} \theta^{\hat{a}}_\mu -
\omega^{\hat{c}}\, _{\hat{a}\mu} \theta^{\hat{a}}_\nu = 0. \label{eq:11}
\end{equation}
Equation (\ref{eq:10}) is formally identical to that for Yang Mills curvature $\{F^{\hat{a}}_{\hphantom{\hat{a}}\hat{c}\mu\nu} dx^\mu \wedge dx^\nu\}$  in terms of its connection $\{A^{\hat{a}}_{\hphantom{\hat{a}}\hat{c}\mu} dx^\mu\}$ but Eq. (\ref{eq:11}) has no correspondent in Yang-Mills theory wherein the connection is the fundamental field.

	The Ricci tensor also has no analogue in Yang-Mills theory but when the contracted Bianchi identities are combined with the vanishing Ricci tensor (vacuum field equation) condition they imply the vanishing of the divergence of spacetime curvature which is an equation of precisely Yang-Mills type. Furthermore, in each case one can compute the divergence of the associated Bianchi identity, commute covariant derivatives and impose the vanishing of the divergence of curvature to derive a natural hyperbolic equation satisfied by the corresponding curvature tensor. For the Einstein problem, expressed in the Cartan formalism, this wave equation for curvature takes the form:
\begin{equation}\label{eq:12}
\begin{split}
\nabla^\alpha\nabla_\alpha R^{\hat{a}}\,_{\hat{b}\mu\nu} +
R_{\mu\nu}\,^{\rho\sigma} R^{\hat{a}}\,_{\hat{b}\rho\sigma} & = 2R^{\hat{a}}\,_{\hat{c}\mu\sigma} R^{\hat{c}}\,_{\hat{b}\nu} ~
^\sigma - 2R^{\hat{a}}\,_{\hat{c}\nu\sigma} R^{\hat{c}}\,_{\hat{b}\mu}
~ ^\sigma \\
&- g^{\alpha\beta} \{ \nabla_\beta [\omega^{\hat{a}}\,_{\hat{c}\alpha}
R^{\hat{c}}\,_{\hat{b}\mu\nu} - R^{\hat{a}}\,_{\hat{c}\mu\nu}
\omega^{\hat{c}}\,_{\hat{b}\alpha} ]  \\
&+ \omega^{\hat{a}}\,_{\hat{c}\beta} [\nabla_\alpha
R^{\hat{c}}\,_{\hat{b}\mu\nu} + \omega^{\hat{c}}\,_{\hat{d}\alpha}
R^{\hat{d}}\,_{\hat{b}\mu\nu} - R^{\hat{c}}\,_{\hat{d}\mu\nu}
\omega^{\hat{d}}\,_{\hat{b}\alpha} ] \\
&- [\nabla_\alpha R^{\hat{a}}\,_{\hat{c}\mu\nu} +
\omega^{\hat{a}}\,_{\hat{d}\alpha} R^{\hat{d}}\,_{\hat{c}\mu\nu} -
R^{\hat{a}}\,_{\hat{d}\mu\nu} \omega^{\hat{d}}\,_{\hat{c}\alpha}
]\omega^{\hat{c}}\,_{\hat{b}\beta} \}
\end{split}
\end{equation}
where here, $\nabla_\alpha$ designates the covariant derivative with respect to spacetime indices only, which ignores frame indices, and the `correction' terms for the latter are reinstated explicitly through the terms involving $\omega^{\hat{a}}_{\hphantom{\hat{a}}\hat{c}\mu}$ that have been moved over to the right.

	The operator acting on curvature on the left hand side of Eq. (\ref{eq:12}) has the same form as that acting on the Faraday tensor of a solution to Maxwell's equations on a vacuum background spacetime. If one pretends for the moment that the terms on the right side of Eq. (\ref{eq:12}) are a given `source' for this Maxwell-like field then it is straightforward to apply the well-known Hadamard/Friedlander analysis of wave equations on curved spacetimes \cite{Hadamard:1923,Friedlander:1975} to write an integral expression for this tensor in terms of integrals over the past light cone from an arbitrary spacetime point p to an `initial', Cauchy hypersurface and additional integrals over the intersection of this cone with the initial surface. Of course for the present problem these `source' terms are not really given since they all involve the unknown but, for nonlinear problems generally, wherein one could hardly expect to derive a true \textit{representation} formula for the solution, this analysis will nevertheless yield an integral equation that can serve as the basis for making \textit{light-cone estimates} of the unknown.

	In a curved spacetime however, where Huygens' principle fails to hold in general, the resulting Hadamard/Friedlander formulas are complicated by the appearance of integrals not only over the (3-dimensional) mantles of the light cones in question and their (2-dimensional) intersections with the initial, Cauchy surfaces but also by integrals over the (4-dimensional) interiors of those cones and their (3-dimensional) intersections with the initial hypersurfaces. It has recently been realized however that one can transform the conventional Hadamard/Friedlander formulas in such a way that only certain integrals over the 3-dimensional mantles of the cones involved and their 2-dimensional intersections with the initial, Cauchy surfaces actually occur \cite{Moncrief:2005,Moncrief:inprep}. At first sight it might seem that one has thereby miraculously restored Huygens' principle even in a curved spacetime where one knows it shouldn't hold, but this is not the case. For purely linear wave equations for example (for which the meaning of Huygens' principle is transparent) this procedure invariably produces integrals over the cone mantles that involve the unknowns themselves in contrast to the original Hadamard/Friedlander formulation which provides  genuine, explicit representation formulas for the solutions of \textit{linear} equations in terms of their Cauchy data (albeit ones with the aforementioned Huygens' principle violating complications).

	The Hadamard/Friedlander formulas are most conveniently expressed in terms of normal coordinates $\{x^\nu\}$ based at the vertex of the light cone in question and defined throughout a normal neighborhood of this point \cite{Moncrief:2005,Moncrief:inprep,LeFloch:2007}. When the Cartan formalism is employed one can most naturally fix the associated orthonormal frame (throughout such a normal neighborhood in terms of its arbitrarily chosen value at the vertex point) by a parallel propagation condition (analogous to the so-called Cronstr\"{o}m condition often used with the Yang-Mills equations \cite{Eardley:1982,Eardley:1982b,Crusciel:1997,Klainerman:2007}) that takes the form:
\begin{equation}
< \omega^{\hat{c}}\,_{\hat{a}}, \tilde{v} > =
\omega^{\hat{c}}\,_{\hat{a}\nu}x^{\nu} = 0. \label{eq:13}
\end{equation}
Remarkably, in this gauge one can not only compute the connection explicitly in terms of curvature (as  Cronstr\"{o}m showed for the Yang-Mills problem) via
\begin{equation}
\omega^{\hat{c}}\, _{\hat{a}\mu} (x) = - \int^1_0 d\lambda ~~ \lambda
x^\nu R^{\hat{c}} \, _{\hat{a}\mu\nu} (x\cdot\lambda), \label{eq:14}
\end{equation}
but also express the orthonormal (co-) frame in terms of the connection (and hence the curvature) by
\begin{equation}\label{eq:15}
\theta^{\hat{c}}_\mu (x) = \theta^{\hat{c}}_\mu (0)  + \int^1_0 d\lambda [\omega^{\hat{c}}\,_{\hat{a}\mu} (\lambda
x)(\lambda x^\nu \theta^{\hat{a}}_\nu (0))].
\end{equation}
	When the aforementioned reduction transformation is applied to the wave equation for spacetime curvature itself, the resulting integral equation may be expressed as:
\begin{equation}\label{eq:16}
\begin{split}
R_{\hphantom{\hat{a}}\hat{b}\alpha\beta}^{\hat{a}} (x) &= \theta_{\alpha}^{\hat{e}} (x) \theta_{\beta}^{\hat{f}} (x) \left\lbrace\frac{1}{2\pi} \int_{C_{p}} \mu_{\Gamma} (x') \left\lbrace\left\lbrack -\omega_{\hphantom{\hat{d}}\hat{e}\sigma'}^{\hat{d}} (x') D^{\sigma'} \left(\kappa (x,x') R_{\hphantom{\hat{a}}\hat{b}\hat{d}\hat{f}}^{\hat{a}} (x')\right)\right.\right.\right.\\
& - \omega_{\hphantom{\hat{d}}\hat{f}\sigma'}^{\hat{d}} (x') D^{\sigma'} \left(\kappa (x,x') R_{\hphantom{\hat{a}}\hat{b}\hat{e}\hat{d}}^{\hat{a}} (x')\right) - \omega_{\hphantom{\hat{d}}\hat{b}\sigma'}^{\hat{d}} (x') D^{\sigma'} \left(\kappa (x,x') R_{\hphantom{\hat{a}}\hat{d}\hat{e}\hat{f}}^{\hat{a}} (x')\right)\\
& \left. + \omega_{\hphantom{\hat{a}}\hat{d}\sigma'}^{\hat{a}} (x') D^{\sigma'} \left(\kappa (x,x') R_{\hphantom{\hat{d}}\hat{b}\hat{e}\hat{f}}^{\hat{d}} (x')\right)\right\rbrack\\
& + \kappa (x,x') \left\lbrack - 2 R_{\hphantom{\hat{a}}\hat{c}\hat{e}\hat{d}}^{\hat{a}} (x') R_{\hphantom{\hat{c}}\hat{b}\hat{f}}^{\hat{c}\hphantom{\hat{b}\hat{f}}\hat{d}} (x') + 2 R_{\hphantom{\hat{a}}\hat{c}\hat{f}\hat{d}}^{\hat{a}} (x') R_{\hphantom{\hat{c}}\hat{b}\hat{e}}^{\hat{c}\hphantom{\hat{b}\hat{e}}\hat{d}} (x')\right.\\
& \left. + R_{\hphantom{\hat{a}}\hat{b}\hat{c}\hat{d}}^{\hat{a}} (x') R_{\hat{e}\hat{f}}^{\hphantom{\hat{e}\hat{f}}\hat{c}\hat{d}} (x')\right\rbrack\\
& + R_{\hphantom{\hat{a}}\hat{b}\hat{e}\hat{f}}^{\hat{a}} (x') \left(\nabla^{\gamma'}\nabla_{\gamma'} \kappa (x,x')\right)\\
& + \left( 2\nabla^{\sigma'} \kappa (x,x')\right) \cdot \left\lbrack\omega_{\hphantom{\hat{d}}\hat{e}\sigma'}^{\hat{d}} (x') R_{\hphantom{\hat{a}}\hat{b}\hat{d}\hat{f}}^{\hat{a}} (x') + \omega_{\hphantom{\hat{d}}\hat{f}\sigma'}^{\hat{d}} (x') R_{\hphantom{\hat{a}}\hat{b}\hat{e}\hat{d}}^{\hat{a}} (x')\right.\\
& \left.\left.+ \omega_{\hphantom{\hat{c}}\hat{b}\sigma'}^{\hat{c}} (x') R_{\hphantom{\hat{a}}\hat{c}\hat{e}\hat{f}}^{\hat{a}} (x') - \omega_{\hphantom{\hat{a}}\hat{c}\sigma'}^{\hat{a}} (x') R_{\hphantom{\hat{c}}\hat{b}\hat{e}\hat{f}}^{\hat{c}} (x')\right\rbrack\right\rbrace\\
& + \frac{1}{2\pi} \int_{\sigma_{p}} d\sigma_{p} \left\lbrace 2\kappa (x,x') \left(\xi^{\sigma'} (x') D_{\sigma'} R_{\hphantom{\hat{a}}\hat{b}\hat{e}\hat{f}}^{\hat{a}} (x')\right)\right.\\
& + \kappa (x,x') \Theta (x') R_{\hphantom{\hat{a}}\hat{b}\hat{e}\hat{f}}^{\hat{a}} (x')\\
& + \kappa (x,x') \xi^{\sigma'} (x') \left\lbrack R_{\hphantom{\hat{a}}\hat{b}\hat{d}\hat{f}}^{\hat{a}} (x') \omega_{\hphantom{\hat{d}}\hat{e}\sigma'}^{\hat{d}} (x')\right.\\
& + R_{\hphantom{\hat{a}}\hat{b}\hat{e}\hat{d}}^{\hat{a}} (x') \omega_{\hphantom{\hat{d}}\hat{f}\sigma'}^{\hat{d}} (x') + R_{\hphantom{\hat{a}}\hat{d}\hat{e}\hat{f}}^{\hat{a}} (x') \omega_{\hphantom{\hat{d}}\hat{b}\sigma'}^{\hat{d}} (x')\\
& \left.\left.\left. - R_{\hphantom{\hat{d}}\hat{b}\hat{e}\hat{f}}^{\hat{d}} (x') \omega_{\hphantom{\hat{a}}\hat{d}\sigma'}^{\hat{a}} (x')\right\rbrack\right\rbrace\vphantom{\frac{1}{2\pi}}\right\rbrace
\end{split}
\end{equation}
where the notation follows that of \cite{Moncrief:2005,Moncrief:inprep} which in turn is based on that of Friedlander \cite{Friedlander:1975}. As promised, only integrals over the light cone mantle $C_p$ and over its (two-dimensional) intersection $\sigma_p$ with the initial surface now occur. One can, by a further transformation, trade the derivatives of the curvature appearing in the light cone integrals above for terms involving the divergence of the Lorentz connection which seems, superficially at least, to be an improvement \cite{Yang-Mills}. But this latter formulation has always proven more problematic to estimate (even in the corresponding Yang-Mills case) than the former one so we shall here sketch what seems to be the most promising approach.

	By well known methods, which have their origins in the original studies of the Yang-Mills problem \cite{Eardley:1982,Eardley:1982b,Crusciel:1997,Klainerman:2007}, one can bound the integrals of those terms that are purely algebraic in curvature by expressions that involve the fluxes of the Bel-Robinson energy. The latter would be controlled by the Bel-Robinson energy but, unlike in the Yang-Mills problem, this natural energy is not itself, a priori under control. Of course the Bel-Robinson energy would be strictly conserved in the presence of a (conformal) Killing field but the existence of such a field is an absurdly strong restriction to place on spacetimes of interest.

	However, when the orthonormal frame fields of the Cartan formalism are subjected (without loss of generality) to the parallel propagation gauge fixing condition described above one can show that these fields (when parallel propagated from the vertex of a particular light-cone) satisfy the Killing equations approximately, with an error term that is explicitly expressible in terms of curvature and that tends to zero at a well defined rate as one approaches the vertex of the chosen cone \cite{Moncrief:2005,Moncrief:inprep}:
\begin{equation}
\theta_{\mu ;\nu}^{\hat{a}} + \theta_{\nu ;\mu}^{\hat{a}} = -
\omega^{\hat{a}}\,\,_{\hat{b}\nu} \theta_{\mu}^{\hat{b}} -
\omega^{\hat{a}}\,\,_{\hat{b}\mu} \theta_{\nu}^{\hat{b}}. \label{eq:17}
\end{equation}
	To handle the terms in Eq. (\ref{eq:16}) involving the (covariant) gradients of curvature one needs a higher order energy for curvature and the expression for this that seems most natural from the point of view taken herein is provided by:
\begin{equation}\label{eq:18}
\tilde{T}_{\mu\nu}^{\mathrm{grav}} := D_{\mu} R \cdot D_{\nu} R - \frac{1}{2} g_{\mu\nu} D_{\gamma} R \cdot D^{\gamma} R
\end{equation}
where now
\begin{equation}\label{eq:19}
\begin{split}
D_{\mu} R^{\hat{a}}_{\hphantom{\hat{a}}\hat{b}\hat{e}\hat{f}} &= \partial_{\mu} R^{\hat{a}}_{\hphantom{\hat{a}}\hat{b}\hat{e}\hat{f}} \\
&\> + R^{\hat{c}}_{\hphantom{\hat{c}}\hat{b}\hat{e}\hat{f}} \omega^{\hat{a}}_{\hphantom{\hat{a}}\hat{c}\mu} - R^{\hat{a}}_{\hphantom{\hat{a}}\hat{c}\hat{e}\hat{f}} \omega^{\hat{c}}_{\hphantom{\hat{c}}\hat{b}\mu}\\
&\> - R^{\hat{a}}_{\hphantom{\hat{a}}\hat{b}\hat{c}\hat{f}} \omega^{\hat{c}}_{\hphantom{\hat{c}}\hat{e}\mu} - R^{\hat{a}}_{\hphantom{\hat{a}}\hat{b}\hat{e}\hat{c}} \omega^{\hat{c}}_{\hphantom{\hat{c}}\hat{f}\mu}
\end{split}
\end{equation}
with
\begin{equation}\label{eq:19a}
R \cdot R = \sum_{\hat{a},\hat{b},\hat{e}\hat{f}} (R^{\hat{a}}_{\hphantom{\hat{a}}\hat{b}\hat{e}\hat{f}})^{2}.
\end{equation}

	The analogous derivations can all be applied to the Yang-Mills problem and shown to yield a dramatically simplified proof of the no-blow-up of Yang-Mills curvature on a curved (globally hyperbolic background) \cite{Moncrief:inprep} but of course the Yang-Mills problem is significantly less challenging than the gravitational one in that, for Yang-Mills fields, the orthonormal frame field and it's (spacetime) curvature are part of the given background and do not require control. How best to modify the arguments in the Einstein problem to achieve the optimal results is currently under intense investigation.

	It should be especially interesting to develop these techniques further and to use them to study the vital interplay between spatial topology and global evolution. Do any \textit{spherical space form} or \textit{handle} summands in the prime decomposition always tend to recollapse and `pinch off' from the $K(\pi, 1)$ summands even as the model universe as a whole continues to expand? If so would some kind of mathematical surgery be necessary (as it is in Ricci flow) to allow the evolution to continue and, if so, what implications does this have for the existence, or perhaps non-existence, of such spherical factors in the actual universe? Do the \textit{graph manifold} components, though continuing to expand always play a comparatively negligible role, through collapse of their \textit{rescaled} metrics, in the asymptotic evolutions? Are the Cauchy hypersurfaces always asymptotically volume-dominated by their \textit{hyperbolic} components with the rescaled metrics on these components asymptotically approaching homogeniety and isotropy? How is the fundamental question of \textit{cosmic censorship} influenced by answers to these questions?

\section*{Acknowledgement}
Moncrief is grateful to Robert Bartnik for making possible his two, very enriching professional visits to Australia and for conveying to him numerous valuable insights into the properties of CMC slicings of Einsteinian spacetimes. This article is dedicated to Bartnik on the occasion of his 60th birthday.

\appendix
\section{Local Well-Posedness of the Einstein-$\Lambda$ Cauchy Problem}
\label{app:local}
Andersson and Moncrief proved in Ref.~\cite{Andersson:2003} a well-posedness theorem for the Cauchy problem for a family of elliptic-hyperbolic systems that included the  (\(n+1\)--dimensional) vacuum Einstein equations in CMSCH gauge. We shall sketch herein how to apply their theorem to the gauge-fixed Einstein-\(\Lambda\) field equations given by (\ref{eq:114}--\ref{eq:115}) and (\ref{eq:107}). Since the Einstein-\(\Lambda\) field equations only differ from the vacuum equations by the addition of some rather innocuous, low order terms most of the technicalities of this extended application of their theorem are straightforward to verify. For this reason we shall mostly refer the reader to the relevant sections of \cite{Andersson:2003} rather than reiterate the detailed arguments herein.

There is however a subtle point in the elliptic analysis of this earlier work that deserves a more substantial discussion. It was assumed in \cite{Andersson:2003} that the spatial manifold \textit{M} admitted a Riemannian, `background' metric \(\hat{g}\) of negative sectional curvature to serve as the reference metric for the spatial harmonic gauge condition imposed throughout. This assumption, though not necessary, was sufficient to ensure that the elliptic equation determining the shift field \textit{X} always had a unique solution or, equivalently, that the associated linear operator always defined an isomorphism between the relevant (Sobolev) spaces (c.f., Lemma 5.2 of \cite{Andersson:2003}). While this implicit topological restriction upon the choice of \textit{M} was not unduly limiting for the applications that the authors of \cite{Andersson:2003} had in mind at the time (c.f., their followup articles \cite{Andersson:2011,Andersson:2004}), it could potentially be so for our purposes and thus we should like to remove it from the hypotheses. Fortunately this issue is not a very `delicate' one and we shall be able to replace the aforementioned constraint upon the choice of the `reference' metric \(\hat{g}\) with a different condition that is not, in itself, topologically restrictive.

First of all though we shall discuss the modifications to the arguments of Section 3 of Ref.~\cite{Andersson:2003} that are sufficient to allow their application to the Einstein-\(\Lambda\) field equations when \(\Lambda > 0\) and where \textit{M} is an arbitrary (compact, connected, orientable and smooth) \textit{n}-manifold of negative Yamabe type.

As we have shown already shown in Section~\ref{sec:reduced-Hamiltonian-monotone-decay} above the Lichnerowicz equation (i.e., Hamiltonian constraint \(\mathcal{H}(g,\pi) = 0\)) has a unique, smooth positive solution for the relevant `conformal factor' if and only if the (spatially constant) mean curvature \(\tau\) satisfies the inequality \(\tau^2 > 2n\Lambda/(n-1)\) and has no solutions otherwise. When this holds the linear elliptic operator occurring in the lapse  equation (\ref{eq:218}) (designed to preserve the CMC slicing condition) has trivial kernel and thus defines an isomorphism between the relevant function spaces. Indeed this equation is identical in form to that dealt with in \cite{Andersson:2003} since it results simply from the replacement of one positive constant, \(\tau^2/n\), by another (namely \(\frac{\tau^2}{n} - \frac{2\Lambda}{(n-1)}\) subject to the inequality discussed above). Since the analysis of this lapse equation given in \cite{Andersson:2003} did not depend upon the actual \textit{value} of this constant but only upon its positivity, it goes through without modification for the present case.

The elliptic operator arising in the equation for the shift vector field (designated by \textit{P} in Section 3 and 5 of \cite{Andersson:2003}) remains unchanged upon inclusion of a cosmological constant. Thus the analysis for it given in the earlier reference applies equally well here. On the other hand the operator \textit{P} does depend non-trivially upon the chosen reference metric \(\hat{g}\) which thus enters into the question of when \textit{P} defines an isomorphism between the relevant (Sobolev) spaces. We shall return to this question below but for now simply assume that \(\hat{g}\) has been chosen so that \textit{P} has trivial kernel \(\forall\; g\) (the `dynamical' metric) in an open set that includes the `initial data'. In this setup one may need to alter the choice of \(\hat{g}\) `stroboscopically' as \textit{g} evolves so as to avoid the development of a non-trivial kernel for \textit{P} (which depends upon both metrics). Thus for the sake of topological generality we may need to employ different choices for the (time independent) metric \(\hat{g}\) over different subintervals of the full time of existence (designated \(T^\ast\) in \cite{Andersson:2003}) of any given solution to the field equations but this merely amounts to representing the spacetime being developed in a collection of (overlapping) SH coordinate gauges rather than a single one.

The only modification to the evolution equations needed to accommodate the inclusion of the cosmological constant \(\Lambda\) is the  replacement of \(F_{ij}\) (c.f., Eqs.~(3.12) and (3.13) of Ref.~\cite{Andersson:2003}) by \(F_{ij} + \left(4\Lambda N/(n-1)\right) u_{ij}\) where, in the notation of that reference, we write \((u_{ij}, v_{ij})\) for the symmetric 2-tensors \((g_{ij}, -2k_{ij})\). But the addition of this \textit{linear, algebraic} term to the \(v_{ij}\) evolution equation causes no difficulty in the verification that this \(\Lambda\)-modified system is elliptic-hyperbolic in the sense defined therein. In particular one thus readily proves that the analogue of the Theorem 3.1 of \cite{Andersson:2003} holds for the \(\Lambda\)-modified system of interest to us here.

To show that the constraints and gauge conditions are preserved in time by the \(\Lambda\)-modified evolutionary system we define the quantities
\begin{align}
A &= \operatorname{tr}_g k - t\label{eq:a01}\\
V^k &= g^{ij} \left(\Gamma^k_{ij}(g) - \hat{\Gamma}_{ij}^k(\hat{g})\right)\label{eq:a02}\\
F &= R (g) + (\operatorname{tr}_g k)^2 - |k|^2 - \nabla_i V^i - 2\Lambda\label{eq:a03}\\
D_i &= \nabla_i \operatorname{tr}_g k - 2\nabla^m k_{mi}\label{eq:a04}\\
\intertext{and}
\alpha_{ij} &= \frac{1}{2} (\nabla_i V_j + \nabla_j V_i)\label{eq:a05}
\end{align}
and show, by direct calculation using the \(\Lambda\)-modified evolution equations (\ref{eq:114}--\ref{eq:115}), that the set of constraint and gauge quantities \((A, F, V^i, D_i)\) satisfies exactly the same induced evolution equations as those given in Eqs.~(4.4a--d) of Ref.~\cite{Andersson:2003}. Thus the energy argument given in Section 4 of this reference goes through exactly as before and shows that if \((A, F, V^i, D_i) = 0\) for the initial data \((g^0, k^0)\), then \((A, F, V^i, D_i) \equiv 0\) along the solution curve \((g, k, N, X)\). Thus, in particular, one arrives at the conclusion that the analogue of Theorem 4.2 of \cite{Andersson:2003} holds for the Einstein-\(\Lambda\) field equations discussed herein and thus that each such solution curve defines a spacetime metric \(\bar{g}\) that satisfies Eqs.~(\ref{eq:105}).

Returning now to the question of when the elliptic operator \textit{P} arising in the shift vector field equation defines an isomorphism between the relevant Sobolev spaces we first note that any (globally defined) Killing vector field\footnote{For the special case of \(n = 2\) any conformal Killing field would have this property.\label{note04}} of the `dynamical' metric \textit{g} would lie in the kernel of this operator whenever the SH gauge condition, \(V^i = 0\), is satisfied. Furthermore such a Killing field (or conformal Killing field in 2 dimensions) would not automatically be \(L^2\)-orthogonal to the `source' term occurring in the shift equation\footnote{Except in extremely special cases such as if \(K_{ij} = \mathrm{constant} \times g_{ij}\).\label{note05}} and thus would provide an immediate obstruction to the \textit{existence} of a solution. Even in those special cases for which a solution did exist the non-trivial kernel of \textit{P} would mitigate against its uniqueness unless a further condition (such as requiring \(L^2\)-orthogonality to the kernel) were imposed.

Fortunately though these aforementioned obstructions are essentially \textit{non-existent} for the `exotic' 3-manifolds of primary interest to us here since one can prove that such manifolds do not admit any Riemannian metric whatsoever that has non-trivial, global Killing symmetries (i.e., that admits a non-vanishing, globally defined Killing field). Any compact manifold that supports a metric with a continuous isometry group (which of necessity would be a compact Lie group) must necessarily admit an SO(2) action and the (compact, connected, orientable) 3-manifolds that do admit such actions have been fully classified.\footnote{In the \(n = 2\) case of (compact, connected, orientable) surfaces one can first `uniformize' an arbitrary metric to one of constant curvature by a suitable conformal transformation and then compute the covariant divergence of the (conformal) Killing equation to derive  an elliptic equation for the hypothetical (conformal) Killing field from which it readily follows that only the sphere, \(S^2\), and torus, \(T^2\), can support metrics with Killing or conformal Killing symmetrics. The `more exotic' higher genus surfaces cannot.\label{note06}} For a rather complete discussion of this issue we refer the reader to Ref.~\cite{Fischer:1996} and  recall that, in the present context, we only consider those 3-manifolds of \textit{negative Yamabe type}.

To summarize the results discussed therein no such \textit{composite 3-manifold} (i.e., non-trivial connect sum) of negative Yamabe type admits an SO(2) action and the only \textit{prime} such manifolds (again of negative Yamabe type) that can admit such actions are `standalone' \(K(\pi,1)\) manifolds of non-flat type\footnote{The six (compact, connected, orientable) 3-manifolds of flat type are of \textit{zero Yamabe type} and five of these do admit SO(2) actions.\label{note07}} whose fundamental group \(\mbox{\boldmath $\pi$}\) has infinite cyclic center \(\approx Z\) (in which case \(M \rightarrow M/SO(2)\) is a Seifert fibered space determined by it numerical Seifert invariants). Thus any 3-manifold of hyperbolic type or that includes such a (hyperbolic) summand in its prime decomposition (so that the arguments of Section~\ref{sec:integral-spacetime} apply) will not admit any SO(2) action and thus not support \textit{any metric} having non-trivial Killing symmetries. Even for those `exceptional' \(K(\pi,1)\) manifolds listed above the `generic' metric that they do support will have no Killing symmetries.

On the other hand, even for those 3-manifolds of negative Yamabe type that admit \textit{no} SO(2) actions (and hence no metrics having non-trivial Killing symmetries) there is the remaining possibility (that we have not excluded) that the operator \textit{P} could still have a non-trivial kernel even when no Killing fields are present to provide one `automatically'. The kernel of such an elliptic operator, however, must necessarily be \textit{finite dimensional} and one can now exploit the non-trivial dependence of \textit{P} upon the `arbitrary' reference metric \(\hat{g}\) to ensure that this kernel vanishes for all \textit{g} in an open neighborhood of the initial data \(g^0\).\footnote{Note that this flexibility is not present when Killing fields occur since any such field will automatically lie in the kernel of \textit{P} independently of the choice of \(\hat{g}\) provided that \(V^i = 0\), as we have assumed.\label{note08}} If this is accomplished (by a suitable choice of \(\hat{g}\)) then one can evolve the initial data for some (open) interval of time and then (`stroboscopically') change the choice of \(\hat{g}\) as needed to avoid the development of a non-trivial kernel.

If for example the self-adjoint operator \(Q_g\) (to which \textit{P} reduces if one takes \(g = \hat{g}\)) defined by
\begin{equation}\label{eq:a06}
Q_g Y^i = g^{mn} \nabla_m\nabla_n Y^i + R^i_{\hphantom{i}j} (g) Y^j
\end{equation}
should happen to have trivial kernel at the initial data metric \(g^0\) we could choose \(\hat{g} = g^0\) and hold this metric fixed as \textit{g} evolves until such time as a new choice of \(\hat{g}\) is called for. If however \(Q_{g^0}\) should happen to have non-trivial kernel then one can seek to perturb the choice of \(\hat{g}\) away from \(g^0\) (while maintaining the SH gauge condition \(V^i = 0\)) in such a way as to eliminate the kernel of \textit{P}.

In this regard let us first note that, on the manifolds of interest herein, the equation
\begin{equation}\label{eq:a07}
\begin{split}
V^i &:= g^{mn} \left(\Gamma_{mn}^i(g) - \hat{\Gamma}_{mn}^i(\hat{g})\right)\\
 &= 0
\end{split}
\end{equation}
defines, in a suitable function space setting \cite{Fischer:1979}, a sub-manifold in the space of metrics \(\hat{g}\) for all \(\hat{g}\) sufficiently near the `fixed' metric \textit{g}.\footnote{A virtually identical argument would show that \(V^i = 0\) also defines a submanifold in the space of metrics \textit{g}, holding \(\hat{g}\) fixed, for all \textit{g} sufficiently near \(\hat{g}\).\label{note09}} Computing the differential of \(V^i\) with respect to \(\hat{g}\) and evaluating the result at \(\hat{g} = g\) one gets
\begin{equation}\label{eq:a16}
\left.\vphantom{\frac{1}{2}} D_{\hat{g}} V^k(\hat{g}) \cdot h \right|_{\hat{g} = g} = -g^{k\ell} \left(h_{i\ell}^{\hphantom{i\ell}|i} - \frac{1}{2} (h^{\hphantom{i}i}_i)_{|\ell}\right)
\end{equation}
where here we write \(|\ell\) for \(\nabla_\ell\), covariant differentiation with respect to \textit{g}.
The adjoint operator \(D_{\hat{g}} V^\ast\) is readily found to be
\begin{equation}\label{eq:a08}
\left( D_{\hat{g}} V^\ast (\hat{g}) \cdot \omega\right)_{ij} = \frac{1}{2} \left(\omega_{i|j} + \omega_{i|j} - g_{ij} \omega^\ell_{\hphantom{\ell}|\ell}\right)
\end{equation}
where \(\omega = \omega_i dx^i\) is an arbitrary one-form field. This operator is easily seen to have injective (principal) symbol (and hence to be \textit{elliptic}) and to have \textit{trivial kernel} on the manifolds of interest since any element of the kernel would have to be a Killing field\footnote{Or, conformal Killing field if \(n = 2\) but, as shown above, these are non-existent for higher genus surfaces.\label{note10}} of (\textit{M, g}) and, as we have shown, these are non-existent.

It thus follows from standard arguments (c.f., especially section 4.2 of Ref.~\cite{Fischer:1979}) that the equation \(V^i(\hat{g}) = 0\) defines a submersion at \(\hat{g} = g\) and hence yields a submanifold in the space of metrics \(\hat{g}\) for all \(\hat{g}\) sufficiently near \textit{g}. As mentioned above a corresponding result would hold for all metrics \textit{g} sufficiently near a fixed \(\hat{g}\) since, except for overall signs, the differential \(D_g V^k(g)\) and its adjoint, computed by holding \(\hat{g}\) fixed, are identical to those given above for the case at hand.

The tangent space to the manifold of \(\hat{g}\) metrics defined by \(V^k(\hat{g}) = 0\) can be conveniently computed by exploiting the (non-\(L^2\)-orthogonal ) decomposition of symmetric 2-tensors at \(\hat{g} = g\) given by\footnote{This decomposition can be derived from the closely related (\(L^2\)-orthogonal) one
\begin{equation*}
h_{ij} = h^{TT}_{ij} + \frac{1}{n} \chi_{g_{ij}} + \nabla_k Z_j + \nabla_j Z_i - \frac{2}{n} g_{ij} \nabla_\ell Z^\ell
\end{equation*}
by simply setting \(\psi = \chi - 2g_{ij} \nabla_\ell Z^\ell\). If (\textit{M, g}) admits conformal Killing fields the choice of \(Z^i\) (and hence also \(\psi\)) could be rendered unique by requiring that \(Z^i\) be \(L^2\)-orthogonal to all of them.\label{note11}}
\begin{equation}\label{eq:a09}
h_{ij} = h^{TT}_{ij} + \frac{1}{n} \psi g_{ij} + \nabla_i Z_j + \nabla_j Z_i
\end{equation}
where \(h^{TT}\) is `transverse-traceless' with respect to \textit{g}, i.e., satisfies
\begin{equation}\label{eq:a10}
g^{ij} h^{TT}_{ij} = 0 \qquad \hbox{ and }\qquad \nabla^j h^{TT}_{ij} = 0.
\end{equation}

To preserve the SH gauge condition \(V^k = 0\) any curve of metrics \(\hat{g}(\lambda)\) with \(\hat{g}(0) = g\) and \(\hat{g}'(0) = h\) would need to satisfy
\begin{equation}\label{eq:a11}
\begin{split}
\left.\frac{d}{d\lambda} V^k\left(\hat{g}(\lambda)\right)\right|_{\lambda = 0} &= \left.\vphantom{\frac{1}{n}}D_{\hat{g}} V^k(\hat{g}) \cdot h\right|_{\hat{g} = g}\\
 &= 0.
\end{split}
\end{equation}
The general solution to this equation for the tangent field \textit{h} is readily found to be
\begin{equation}\label{eq:a12}
h_{ij} = \hat{h}^{TT}_{ij} + \frac{1}{n} \hat{\psi} g_{ij} + \nabla_i \hat{Z}_j(\hat{\psi}) + \nabla_j \hat{Z}_i(\hat{\psi})
\end{equation}
where \(\hat{h}^{TT}\) is an \textit{arbitrary} (transverse-traceless, symmetric) 2-tensor, \(\hat{\psi}\) is any scalar field whose gradient is \(L^2\)-orthogonal to the kernel of \(Q_g\) and where \(\hat{Z}_i(\hat{\psi})\) is a solution to
\begin{equation}\label{eq:a13}
\begin{split}
Q_g\hat{Z}_i &:= g^{mn} \nabla_m\nabla_n\hat{Z}_i + R^j_{\hphantom{j}i}(g)\hat{Z}_j\\
 &= \left(\frac{1}{2} - \frac{1}{n}\right)\hat{\psi}_{|i}
\end{split}
\end{equation}
which we may require to be \(L^2\)-orthogonal to the kernel of \(Q_g\).\footnote{We here relax the aforementioned condition that \(\hat{Z}_i\) be \(L^2\)-orthogonal to any conformal Killing fields of \textit{g}.\label{note12}}

Now suppose that the kernel of \(Q_g\) is spanned by a collection of \textit{k} vector fields \(\left\lbrace\overset{(A)}{Y}|\; A = 1, \ldots , k\right\rbrace\) and ask whether we can smoothly deform these along a chosen smooth curve of metrics \(\hat{g}(\lambda)\) (with \(\hat{g}(0) = g\) and \(\hat{g}'(0) = h\)) that preserves the SH condition \(V^i\left(\hat{g}(\lambda)\right) = 0\) to curves of vector fields
\begin{equation}\label{eq:a14}
\left\lbrace\overset{(A)}{Y}(\lambda)| A = 1, \ldots k\qquad \hbox{ with }\qquad \overset{(A)}{Y}(0) = \overset{(A)}{Y}\right\rbrace
\end{equation}
that, either all or in part, continue to lie in the kernel of the operator \textit{P} which, through its dependence upon \(\hat{g}\) now varies smoothly with \(\lambda\). Computing the derivative of the defining condition, \(P_\lambda \overset{(A)}{Y}(\lambda) = 0\), with respect to \(\lambda\) and setting \(\lambda = 0\) one readily finds that a necessary condition for \(\overset{(A)}{Y}(\lambda)\) to remain in the kernel of \(P_\lambda\) is that
\begin{equation}\label{eq:a15}
\int_M d\mu_g\; \left\lbrace\overset{(B)}{Y_i} \left(\nabla^m \overset{(A)}{Y^n} + \nabla^n \overset{(A)}{Y^m}\right) \left.D_{\hat{g}} \hat{\Gamma}^i_{mn}(\hat{g}) \cdot h\right|_{\hat{g} = g}\right\rbrace = 0
\end{equation}
for \(B = 1, \ldots k\). In the language of Rayleigh-Schr\"{o}dinger perturbation theory familiar from quantum mechanics the above correspond to the `matrix elements' of the perturbation (to the operator \textit{P}) that must necessarily vanish if the vanishing of the corresponding eigenvalue is to be preserved.

The main point is that, for a `generic' choice of the perturbation \textit{h}, one expects the conditions (\ref{eq:a15}) to be maximally violated and thus for the kernel of \textit{P} to be fully annihilated by translation of \(\hat{g}\) along the corresponding curve of metrics \(\hat{g}(\lambda)\). While a fully rigorous treatment of this issue would take some further work we remind the reader that the starting point of this analysis was the assumption that the operator \(Q_g\) had a non-trivial kernel which, itself, is seemingly a highly non-generic restriction on the initial metric \textit{g}.

{}
\end{document}